\pdfoutput=1 
\documentclass[a4paper,11pt]{article}

\usepackage{jheppub}
\usepackage[utf8]{inputenc}
\usepackage[colorlinks=true,citecolor=blue,linkcolor=blue]{hyperref}
\usepackage{amsmath}
\usepackage{amssymb}
\usepackage{soul,xcolor}
\usepackage{epsfig}
\usepackage{epstopdf}
\usepackage{url}
\usepackage{float}
\usepackage{slashed}
\usepackage{cancel}
\usepackage{textcomp}
\usepackage{calc}
\usepackage{adjustbox}
\setstcolor{red}
\usepackage[T1]{fontenc} % if needed

\allowdisplaybreaks

\catcode`\@=11
\def\lsim{\mathrel{\mathpalette\@versim<}}
\def\gsim{\mathrel{\mathpalette\@versim>}}
\def\@versim#1#2{\vcenter{\offinterlineskip
\ialign{$\m@th#1\hfil##\hfil$\crcr#2\crcr\sim\crcr } }}
\catcode`\@=12
\newcommand{\e}{{\rm e}}
\newcommand{\al}[1]{\begin{align}#1\end{align}}
\newcommand{\nn}{\nonumber}

\newcommand{\GeV}{\,\mathrm{GeV}}

\newcommand{\be}{\begin{eqnarray}}
\newcommand{\ee}{\end{eqnarray}}

\title{\boldmath Unified Emergence of Energy Scales and Cosmic Inflation}

\author[a,b]{Jisuke Kubo,}
\author[a]{Jeffrey Kuntz,}
\author[a]{Manfred Lindner,}
\author[a]{Jonas Rezacek,}
\author[a]{Philipp Saake,}
\author[a]{Andreas Trautner}

\affiliation[a]{Max-Planck-Institut f\"ur Kernphysik (MPIK),\\Saupfercheckweg 1, 69117 Heidelberg, Germany}
\affiliation[b]{Department of Physics, University of Toyama,\\3190 Gofuku, Toyama 930-8555, Japan}

\emailAdd{kubo@mpi-hd.mpg.de}
\emailAdd{jeffrey.kuntz@mpi-hd.mpg.de}
\emailAdd{lindner@mpi-hd.mpg.de}
\emailAdd{jonas.rezacek@mpi-hd.mpg.de}
\emailAdd{philipp.saake@mpi-hd.mpg.de}
\emailAdd{trautner@mpi-hd.mpg.de}

\abstract{In the quest for unification of the Standard Model with gravity, classical scale invariance can be utilized to dynamically generate the Planck mass $M_\mathrm{Pl}$.
However, the relation of Planck scale physics to the scale of electroweak symmetry breaking $\mu_H$ requires further explanation. 
In this paper, we propose a model that uses the spontaneous breaking of scale invariance in the scalar sector as a unified origin for dynamical generation of both scales. 
Using the Gildener-Weinberg approximation, only one scalar acquires a vacuum expectation value of $v_S \sim (10^{16-17}) \GeV$, thus radiatively generating $M_\mathrm{Pl} \approx \beta_S^{1/2} v_S$ 
and $\mu_H$ via the neutrino option with right handed neutrino masses $m_N = y_M v_S \sim 10^7 \GeV$. 
Consequently, active SM neutrinos are given a mass with the inclusion of a type-I seesaw mechanism. 
Furthermore, we adopt an unbroken $Z_2$ symmetry and a $Z_2$-odd set of right-handed Majorana neutrinos $\chi$ that do not take part in the neutrino option 
and are able to produce the correct dark matter relic abundance (dominantly) via inflaton decay. 
The model also describes cosmic inflation and the inflationary CMB observables 
are predicted to interpolate between those of $R^2$ and linear chaotic 
inflationary model and are thus well within the strongest experimental constraints.}

\begin{document} 
\maketitle
\flushbottom

\section{Introduction}

The quest to dynamically generate the Planck mass 
$m_\mathrm{Pl}$ (or the reduced Planck mass $M_\mathrm{Pl}=m_\mathrm{Pl}/\sqrt{8 \pi}$)
has existed for quite some time. 
Scale invariance, whether it be global~\cite{Fujii:1974bq,Minkowski:1977aj,Zee:1978wi,Terazawa:1981ga}
(and recently in Refs.~ \cite{Salvio:2014soa, Kannike:2015apa, Kannike:2016wuy, Einhorn:2014gfa, Donoghue:2018izj, Donoghue_2018,Holdom_2016,Hill:2018qcb, Ferreira:2016wem,Vicentini:2019etr})
or local~\cite{Englert:1975wj,Englert:1976ep,Chudnovsky:1976zj,Fradkin:1978yw, Smolin:1979uz,Zee:1979hy,Nieh:1982nb,Terazawa:1976eq,Akama:1977nw,Akama:1977hr,Adler:1980bx,Adler:1980pg,Zee:1980sj,Adler:1982ri} 
(recently in Refs.~\cite{Mannheim:2011ds, Ghilencea:2018dqd,Oda:2019pid,Barnaveli:2018dxo}),
scale symmetry has played a central role
based on the fact that it forbids the presence of an Einstein-Hilbert term in the action.
Similarly to Einstein's theory of gravity, which contains a
single dimensionful parameter $m_\mathrm{Pl}$ (apart from the 
cosmological constant),
the Higgs mass term parameter $\mu_H$
is the only dimensionful parameter 
in the Standard Model (SM) of particle physics.
The scale invariant limit of the SM i.\e.\ in the limit $\mu_H$ goes to zero, was investigated by
Coleman and Weinberg \cite{Coleman:1973jx}, who
found that radiative corrections
can change the tree level form of  the Higgs potential
and thus break the electroweak gauge symmetry spontaneously.
Recent experimental observations in astrophysics and particle physics
indeed hint that 
Einstein's theory and the SM
should both be extended in a (classically)
scale invariant fashion. 

With respect to gravity, the Planck measurements of the CMB \cite{Aghanim:2018eyx,Akrami:2018odb} show that,
not only is the idea of new inflation \cite{Linde:1981mu,Linde:1982zj,Albrecht:1982wi}
consistent with their data, but also that
the scalar spectral index $n_s$ of the gravitational fluctuations
is close to one, and in particular that
the ratio $r$ of tensor to scalar power spectra of 
fluctuations can be nearly zero.
Accordingly, $R^2$ inflation \cite{Starobinsky:1980te,Mukhanov:1981xt,Starobinsky:1983zz} and also Higgs inflation \cite{Bezrukov:2007ep}
seem to be the most promising candidate models
\cite{Akrami:2018odb}. The main reason for the
success is a (super) flat inflationary scalar 
potential expressed in the Einstein frame after a local Weyl scaling from the Jordan frame
\cite{Bezrukov:2007ep,Mijic:1986iv} where
$r$ is proportional to the gradient of the scalar potential. 
The super flatness in the scalar  potential in both models
is caused by the relative suppression of the non scale invariant Einstein-Hilbert term $R$ compared respectively to the scale invariant terms of $R^2$  inflation ($\gamma \,R^2$, with $\gamma \sim O(10^9)$)\cite{Mijic:1986iv,Hwang:2001pu} and Higgs inflation ($\beta |H|^2 R$ with $\beta \,\sim O(10^4)$)\cite{Bezrukov:2007ep}, where $R$ is the Ricci curvature scalar 
and $H$ is the SM Higgs doublet. In this sense, CMB data suggests
a scale invariant extension of Einstein's theory of gravity not only because of the close-to scale invariant spectral tilt $n_s$, but also because of the suppression of $r$.
Furthermore, it can be argued that gravity equipped with local scale symmetry can be rendered renormalizable \cite{Adler:1982ri, tHooft:2011aa,Mannheim:2011ds} 
due to higher derivative terms in the action \cite{Stelle:1976gc}.\footnote{%
The inclusion of the Weyl tensor squared term is well-known to lead to a violation of unitarity. 
This is intimately related to the presence of a spin-2 ghost which is addressed in e.g.\ Refs.~\cite{Bender:2007wu,Donoghue:2019fcb,Anselmi:2018ibi} 
and it is beyond the scope of the present work. Other possible solutions to the ghost problem are based on the presence of higher-curvature terms~\cite{Eliezer:1989cr,Jaen:1986iz,Simon:1990ic,Biswas:2011ar} which would be dangerous for our model as they would modify the scalar potential and 
spoil global scale invariance. Assuming the absence of such terms, hence, is crucial for our discussion but introduces additional model dependence.
}

The SM, on the other hand, describes our microscopic world with remarkable success despite its various shortcomings.
Remarkably, the Higgs mass $m_h$ has turned out to be~\cite{Aad:2012tfa,Chatrchyan:2012ufa} such that the SM remains perturbative 
i.\e.\ it contains no Landau poles, below the Planck scale~\cite{Holthausen:2011aa,Bezrukov:2012sa,Degrassi:2012ry,Buttazzo:2013uya}. That is, the mass parameter $\mu_H$ can logarithmically run
all the way to the Planck scale so that,
according to Bardeen \cite{Bardeen:1995kv},
the SM does not, by itself, have a fine-tuning problem.
We regard this as a strong evidence \cite{Lindner2019} for extending the SM
in a scale invariant way 
\cite{Hempfling:1996ht,Meissner:2006zh}, because the logarithmic running of $\mu_H$ up to
the Planck scale means that scale invariance is broken only by the scale anomaly \cite{Callan:1970yg,Symanzik:1970rt}, except of course, by the soft breaking due to $\mu_H$.

In this paper we thus pursue the idea that 
the Planck mass $m_\mathrm{Pl}$
and the Higgs mass parameter $\mu_H$ have a unified origin,
namely, the spontaneous breaking 
of scale invariance. 
Consequently, a real SM singlet scalar
field $S$ acquires a finite vacuum expectation value~(VEV)~$\langle S \rangle$.
However, since $M_\mathrm{Pl}$ is of order $10^{18}$ GeV
and $\mu_H$ is of order $10^2$ GeV, we must also address
the question of  how it may be possible
to generate $M_\mathrm{Pl}$ and  $\mu_H$ and, in particular, their hierarchy from a common source.
In fact, this is the central question of our scenario, and
even though it may be still far from ultimate and satisfactory, 
our answer demonstrates how to soften this huge hierarchy.

The solution is based on the observation that
heavy right-handed neutrinos $N$ contribute
an important correction to $\mu_H^2$;
the finite term  $\Delta \mu_H^2$ is proportional to
$y_\nu^2 m_N^2/4\pi^2$ 
\cite{Vissani:1997ys,Casas:1999cd,Clarke:2015gwa,Bambhaniya:2016rbb},
where $y_\nu$ stands for the  Dirac-Yukawa coupling
and $m_N$ is the representative mass of $N$.
The large contribution can be used to radiatively generate an appropriately sized $\Delta \mu_H^2$ i.\e.\
$ \Delta \mu_H^2\simeq \mu_H^2$, including its sign.
This is the idea of  the  ``neutrino option" \cite{Brivio:2017dfq},  and
when the seesaw mechanism 
\cite{Minkowski,GellMann:1980vs,Yanagida:1979as,Goran} is implemented
to obtain light active neutrino masses, one finds
that $m_N\sim 10^7$ GeV and $y_\nu \sim 10^{-4}$.
The neutrino option thus establishes a link between
the heavy  right-handed neutrinos and 
the electroweak scale \cite{Brivio:2017dfq}.
This scenario can be neatly extended
in a scale invariant manner \cite{Brdar:2018vjq},
where  the mass of $N$
is generated from the  Majorana-Yukawa coupling
$y_M S N^T C N$, where $C$ is the charge conjugation matrix, hence, $m_N=y_M \langle S \rangle$.

We adopt this mechanism here to soften the huge hierarchy between 
$M_\mathrm{Pl}$ and  $\mu_H$ by many orders of magnitudes. 
The dimensionless parameter $\beta_S$ of the non-minimal coupling $\beta_S S^2 R$ gives $M_\mathrm{Pl}\simeq\sqrt{\beta_S} \,\langle S \rangle$.
Since $\beta_S\sim O(10^{2-3})$ for realistic inflation, we have that $\langle S \rangle\sim O(10^{16-17})$ GeV so that $m_N$ of 
the order $10^7$ GeV can be obtained if $y_M$ is of the order $10^{-(9-10)}$. 
The smallness of $y_M$ is technically natural in the sense of
't Hooft \cite{tHooft:1979rat}, since an anomaly-free global $U(1)_{B-L}$ is restored
in the limit $y_M\rightarrow0$ (together with another Yukawa coupling to be discussed below).
However, the scenario is not without problems:
The Higgs portal coupling $\lambda_{HS}\, S^2\,|H|^2$,
which would give a large contribution to $\mu_H^2$ for large
$\langle S\rangle$, can not be forbidden by any symmetry.
Nonetheless, we can set up the model (at least on a flat background spacetime)
such that the radiative correction to $\lambda_{HS}$
remains sufficiently small such that they do not spoil our scenario of a unified origin of energy scales.
This is possible because the radiative corrections, in the absence of $y_M$, are proportional to $\lambda_{HS}$ itself.

In the next section we begin by writing down the Lagrangian of the model.
The spontaneous scale symmetry breaking is achieved by
the Gildener-Weinberg mechanism \cite{Gildener:1976ih}, for which we introduce
an additional real scalar field $\sigma$.
We also impose a discrete symmetry $Z_2$, which
not only simplifies the form of the scalar potential,
but also stabilizes the $Z_2$-odd particles if it is not spontaneously 
broken. Accordingly, $Z_2$-odd Majorana particles $\chi$ 
are introduced as dark mater candidates.
In section III spontaneous scale symmetry breaking is
discussed, and $M_\mathrm{Pl}$ is identified in such a way that
$m_\mathrm{Pl}$ can be related to the renormalization scale.
In section IV we derive the effective action for inflation
in the Jordan frame and subsequently perform a Weyl transformation to the Einstein frame where
we calculate inflationary parameters.
We explain under which conditions the scalaron-$S$ system can be approximated as a single inflaton system.
We perform benchmark point studies as well as a parameter scan to work out the predictions of the model.
Dark matter is treated in section V. There are two kinds of $Z_2$-odd particles in the model, 
$\sigma$ (boson) and $\chi$ (fermion), and the flat direction approximation of the Gildener-Weinberg mechanism 
works in such a way that the $Z_2$ symmetry remains unbroken. 
Because it is necessary for successful inflation, $\sigma$ becomes heavier than $S$. 
Hence, $S$ can decay into two $\chi$, while a decay into two $\sigma$ is not possible~\cite{Chung:1998rq,Allahverdi:2002nb}.
This process is nothing more than a freeze-in mechanism \cite{Hall:2009bx} for the dark matter $\chi$ so that it may reach the 
observed value of relic abundance. In section VI we briefly review the neutrino option~\cite{Brivio:2017dfq}, and refer to 
the literature cited there for details. The last section is devoted to our conclusions.

\section{The model}

Our total Lagrangian ${\cal L}_\mathrm{T}$ consists of four parts:
(i) ${\cal L}_\mathrm{CW}$ is, after including quantum corrections, responsible for the spontaneous breaking 
of global conformal symmetry and the generation of a unified scale.
(ii) ${\cal L}_\mathrm{GR}$ is responsible for the identification of the Planck scale, thereby 
generating the Einstein-Hilbert action.
(iii) ${\cal L}_\mathrm{SM}$ describes the SM interactions,
and (iv) ${\cal L}_{N\chi}$ is responsible for generating light neutrino masses via
a type-I see-saw mechanism which at the same time radiatively induces the Higgs mass term and accommodates for dark matter.
Altogether, 
\al{
	{\cal L}_\mathrm{T}& ={\cal L}_\mathrm{CW}+{\cal L}_\mathrm{GR}
	+{\cal L}_{{\rm SM}}+{\cal L}_{N\chi} \,,
	\label{FullL1}}
where
\al{
	\frac{{\cal L}_\mathrm{CW} }{\sqrt{-g}}& = \frac{1}{2}g^{\mu\nu}\partial_\mu S \partial_\nu S
	+\frac{1}{2}g^{\mu\nu}\partial_\mu \sigma \partial_\nu \sigma
	-\frac{1}{4}\lambda_S S^4-\frac{1}{4}\lambda_\sigma \sigma^4-
	\frac{1}{4}\lambda_{s\sigma} S^2\sigma^2\,,
	\label{LCW}\\
	\frac{{\cal L}_\mathrm{GR} }{\sqrt{-g}}& = 
	-\frac{ 1}{2}(\beta_S S^2+\beta_\sigma\sigma^2 +  \beta_H H^\dagger H)\,R +\gamma\,R^2+\kappa W_{\mu\nu\alpha\beta} W^{\mu\nu\alpha\beta} \,,
	\label{LCGR}\\
	\frac{{\cal L}_{{\rm SM}} }{\sqrt{-g}} 
	&= \left.{\cal L}_{\rm SM}\right|_{\mu_H=0}-\frac{1}{4}(\lambda_{HS} S^2
	+ \lambda_{H\sigma} \sigma^2)H^\dag H\,,
	\label{LSM}	\\
	\frac{{\cal L}_{N\chi}  }{\sqrt{-g}}&=
	\frac{i}{2} \bar{N}\slashed{\partial} N
	- \frac{1}{2} y_M S N^T C N 	+
	\frac{i}{2} \bar{\chi}\slashed{\partial} \chi
	- \frac{1}{2} y_\chi S \chi^T C \chi 	\nn\\
	& \quad - \left( y_{N\chi }\sigma N^T C \chi 	
	\,+\, y_\nu \bar{L} \tilde{H} \,\tfrac{1}{2}(1+\gamma_5)
	N + \mathrm{h.c.} \right)\,.
	\label{LNchi}
}
Here, $R$ denotes the Ricci curvature scalar, 
$W_{\mu\nu\alpha\beta}$ is the Weyl tensor,
$N$ and $\chi$ denote the (three+three) right-handed  Majorana neutrinos, while
$H$ ($\tilde{H}=i\sigma_2 H^*$) 
and $L$  are the SM Higgs and lepton doublets. $\left.{\cal L}_{\rm SM}\right|_{\mu_H=0}$ is the SM Lagrangian
without the quadratic Higgs term, and we suppress flavor indices throughout, though strictly speaking, the Yukawa couplings 
$y_M, y_\chi, y_{N\chi}$ and $y_\nu$ should all be
matrices in generation space. However, we will not consider details of the flavor structure here, 
and therefore, we treat them as representative real numbers.
The total Lagrangian ${\cal L}_\mathrm{T}$ presents the most general function that respects the SM gauge symmetries, general diffeomorphism invariance,\footnote{
Due to the presence of minimal fermion-gravitational couplings, the use of the vierbein formalism is quietly 
understood with respect to these terms, even though it does not play a role in our analysis.}
global conformal invariance at the classical level, and a discrete $Z_2$ symmetry with $\sigma$ and $\chi$ being the only $Z_2$-odd fields.
We suppress a possible Gau\ss-Bonnet surface term in Eq.~(\ref{LCGR}).

We note that the set of real scalars $S$ and $\sigma$ is the most economic way to successfully realize
the spontaneous breaking of scale invariance \`{a} la Coleman-Weinberg \cite{Coleman:1973jx,Gildener:1976ih}.
We will see that the real scalar $S$ has a triple role:
(i) It is the only scalar that forms a condensate and thereby breaks 
global conformal invariance spontaneously,
(ii) it is the mediator that transmits the energy scale 
inherent in the condensate
to the gravity (${\cal L}_\mathrm{GR}$) and neutrino (${\cal L}_{N\chi}$) sectors, and subsequently to the SM sector, and (iii)
it serves as the inflaton.

\section{Spontaneous conformal breaking and the Planck mass}
 
  In order to simultaneously avoid
  the domain wall problem \cite{Kirzhnits:1972ut} 
  and stabilize the DM candidate
  $\chi$, we choose a flat direction of the scalar potential
  such that the $Z_2$ symmetry remains unbroken.
  The desired (approximate) flat direction,
  $S\neq 0$ and $\sigma=0$, can be  realized if~\cite{Gildener:1976ih}
  \al{
  	\lambda_S & \ll  \lambda_{S\sigma}\quad \mbox{and}\quad\lambda_S \ll\lambda_{\sigma}\;.
  	\label{flatdr}
  }
  As we see from Eq.~(\ref{LCGR}), a non-zero VEV of $S$ denoted by $v_S$
  will generate the Einstein term $ -(1/2)M_\mathrm{Pl}^2\, R$
  with the (reduced) Planck mass $M_\mathrm{Pl}\simeq \sqrt{\beta_S} v_S$.
  At the same time, the Majorana neutrinos $N$ and $\chi$ become massive
  due to the Yukawa interactions in Eq.~(\ref{LNchi}); $m_{N}=y_M \, v_S
  \simeq y_M\,M_\mathrm{Pl}/\sqrt{\beta_S}$ and 
  $m_{\chi}=y_\chi \, v_S
  \simeq y_\chi\,M_\mathrm{Pl}/\sqrt{\beta_S}$.
  Furthermore, in order to utilize the neutrino option we must assume that
  the Higgs portal couplings $\lambda_{HS}$ and $\lambda_{H\sigma}$
  are \textit{extremely} suppressed and that
  $m_N$ is of order $10^7$ GeV, implying that 
  $y_M\sim \sqrt{\beta_S}(m_N/M_\mathrm{Pl})\sim O(10^{-10}) ~
  \mbox{for}~\beta_S\simeq 10^3$.
  It is important to note that the set of couplings 
  $\lambda_{HS}$, $\lambda_{H\sigma}$, $y_M$, $y_\chi$, and $y_{N\chi}$
  remain zero at higher order in perturbation theory if they are set equal to zero at tree-level.
  Therefore, the smallness of these couplings is in some sense natural
  even though no enhancement of symmetry is associated (see, however,~\cite{Foot:2013hna}). 
  Similarly, we assume an approximately vanishing coupling $\beta_H \approx 0$ 
  so that the Higgs-Ricci scalar term in Eq.~(\ref{LCGR}) can be neglected.\footnote{%
  Specifically, 
  we assume that $\beta_H R\ll \lambda_{HS} S^2$ during inflation so that the resulting correction to Eq.~\eqref{Ueff}
  can be neglected. Since $\lambda_{HS}$, $\lambda_{H\sigma}$, and $\beta_H$ are all very small by assumption, the Higgs plays no role in this scenario of inflation.}
  Neglecting the aforementioned couplings we integrate out the quantum fluctuations $\delta S$ and $\delta\sigma$ at one-loop in the background with
  $S\neq 0$ and $\sigma=0$  to obtain the effective potential
  \al{
  	U_\mathrm{eff}(S,R,\sigma)&=   \frac{1}{4}\lambda_S S^4
  	+\frac{1}{4}\lambda_\sigma \sigma^4+
  	\frac{1}{4}\lambda_{s\sigma} S^2\sigma^2+
  	\frac{1}{64 \pi^2}\left(\,
  	\tilde{m}_s^4 \ln  [\tilde{m}_s^2/\mu^2]+
  	\tilde{m}_\sigma^4 \ln  [\tilde{m}_\sigma^2/\mu^2]
  	\,\right)\,,
  	\label{Ueff}}
  where
  \al{\label{ms_msigma}
	\tilde{m}_s^2 &=
  	3 \lambda_S S^2+\beta_S R\,\quad\mathrm{and}\quad\,
  	\tilde{m}_\sigma^2= \frac{1}{2}\lambda_{S\sigma} S^2+\beta_\sigma R \,.
  }
  Here we have used the $\overline{\mbox{MS}}$ scheme and
  the constant $-3/2$ is absorbed into the renormalization scale $\mu$.\footnote{
  The integration does not only give the desired Coleman-Weinberg potential, but also 
  divergences that can be absorbed into $\lambda_S$, $\gamma$, and $ \beta_S$. 
  This agrees with the earlier computation of e.g.\ Ref.~\cite{Casarin:2018odz}~(see also~\cite{Steinwachs:2011zs} and references cited therein).
  Strictly speaking, also note that $\beta_S$ and $\beta_\sigma$ in Eq.~\eqref{ms_msigma} should read $\beta_S-1/6$ and $\beta_\sigma-1/6$, respectively, 
  if one properly takes into account the non-flatness of space-time background and the integration of the quantum fluctuations \cite{Markkanen:2018bfx}. 
  However, since $\beta_S$ will turn out be large (i.e.$\gsim 10^2$) for realistic cosmic inflation and the inflationary parameters will depend barely on $\beta_\sigma$, we will be ignoring the constants $1/6$ throughout the paper.
  }
  
  Because of $\langle \sigma \rangle =0$, the field $\sigma$ does not play any role for inflation and so we suppress it throughout the following discussions.
  To compute  $v_S=\langle S\rangle$, we assume  that $\beta_S  R  < 3 \lambda_S S^2 $ 
  and $\beta_\sigma  R < (1/2) \lambda_{S\sigma} S^2 $ (during inflation), such that
  the effective potential $U_\mathrm{eff}$ in  Eq.~(\ref{Ueff}) can be expanded 
  in powers of 
  $\beta_S R$ and
  $\beta_\sigma  R$,
  \al{
  	U_\mathrm{eff}(S,R,\sigma=0) &=U_0+
  	U_\mathrm{CW}(S)+ U_{(1)}(S) \,R+U_{(2)}(S) R^2 +O(R^3 )\,,
  	\label{Ueff1}
  }
  where 
  \al{  
  	U_\mathrm{CW}(S) &=\frac{1}{4}\lambda_S S^4+\frac{S^4}{64 \pi^2}
  	\left\{ \,9 \lambda_S^2 \ln [3 \lambda_S S^2/\mu^2]+
  	(1/4) \lambda_{S\sigma}^2 \ln [(1/2) \lambda_{S\sigma} S^2/\mu^2]\,\right\}-U_0\,,\label{Ucw}\\
  	U_{(1)}(S) &=\frac{1}{128 \pi^2}
  	\left\{ 6\beta_S  \lambda_S S^2\left(1+2\ln [3 \lambda_S S^2/\mu^2]\right)
  	\nn\right.\\+
  	& \left.\beta_\sigma  \lambda_{S\sigma} S^2\left(1+2\ln [(1/2) \lambda_{S \sigma}S^2/\mu^2]\right)\,\right\}\,,
  	\label{U1}\\
  	U_{(2)}(S) & =\frac{1}{128 \pi^2}
  	\left\{ \beta_S^2  \left(3+2\ln [3 \lambda_S S^2/\mu^2]\right)+
  	\beta_\sigma^2 \left(3+2\ln [(1/2) \lambda_{S \sigma}S^2/\mu^2]\right)\,\right\}\,.
  	\label{U2}
  } 
Since we are assuming a negligibly small (but, of course, nonzero during inflation) value of the curvature scalar $R$, 
we obtain the $R$-independent leading-order $v_S$ from the potential $U_\mathrm{CW}(S)$. 
The zero-point energy density $U_0$ is chosen such that $U_\mathrm{CW}(S=v_S)=0$ is satisfied,
which is consistent with $\langle R \rangle=0$ at the leading order.
That is, our effective potential now reads 
\al{
\tilde{U}_\mathrm{eff}(S,R) = U_\mathrm{eff}(S,R,0)-U_0 = U_\mathrm{CW}(S)+U_{(1)}(S) R+U_{(2)}(S)R^2+O(R^3 )\;,}
and we find that $U_0$ can be written as
\begin{equation}
    U_0 = -\mu^4\,\frac{\beta_{\lambda_S}}{16}\, \exp \left[ -1-16 C/ \beta_{\lambda_S} \right]  \,,
\end{equation}
where
\begin{align}
\beta_{\lambda_S} =\frac{1}{16\pi^2}\left(18\lambda_S^2+  \frac{1}{2}\lambda_{S\sigma}^2\right)\,,\quad \mathrm{and}\quad
C =\frac{1}{4}\lambda_S+\frac{1}{64 \pi^2}\left(\,9 \lambda_S^2 \ln (3\lambda_S)+\frac{1}{4} \lambda_{S\sigma}^2 \ln (\lambda_{S\sigma}/2)\right)\,.
\end{align}
  Note that $\beta_{\lambda_S}$ is the one-loop $\beta$-function
  for  $\lambda_S$ in the absence of $y_M$ and $y_\chi$.
  The negative zero-point energy density $U_0$ is a consequence of
  the spontaneous breaking of global conformal symmetry. 
  This zero-point energy density, which is the cosmological constant,
  is finite in dimensional regularization because of the scale invariance of the total Lagrangian in Eq.~(\ref{FullL1}).
  Our choice to subtract the zero-point energy density corresponds to an explicit super-soft breaking of scale invariance at tree level, which is the cost of remedying the cosmological constant problem in this model.
  Nonetheless, we note that the zero-point energy cannot be uniquely determined within the framework of quantum field theory in flat spacetime. 
  To properly address the cosmological constant problem one should also take into account gravitational quantum fluctuations, including contributions arising from the (possibly) unitarity-violating Weyl tensor term in the action. We set this issue aside for the purpose of this work and continue with our discussion.
  
  The  identification of $M_\mathrm{Pl}$ follows from 
  the first term in Eq.~(\ref{LCGR}) along with Eq.~(\ref{Ueff1}): 
  \al{
  	M_\mathrm{Pl} &= \left(\beta_S +
  	2 U_{(1)}(v_S)/v_S^2 \right)^{1/2} \,v_S\,,
  	\label{mpl}
  }
  where $2 U_{(1)}(v_S)$ can be written in  an analytic form as
  \al{
  	2 U_{(1)}(v_S)=&-\frac{\lambda_S v_S^2}{36\lambda^2_S+\lambda^2_{S\sigma}}
  	\left(\, 12 \beta_S \lambda_S + 2 \beta_\sigma \lambda_{S\sigma}
  	-\frac{3\lambda_{S\sigma}}{16\pi^2}
  	[-6\beta_\sigma \lambda_S +\beta_S \lambda_{S\sigma}]
  	\ln (6 \lambda_S/\lambda_{S\sigma})\right)\,.
  	%\label{mpl}
  }
  Since $v_S=\mu f_S(\lambda_S,\lambda_{S\sigma})$ 
  (as can be seen in Eq.~(\ref{Ucw})), Eq.~(\ref{mpl}) relates
  $M_\mathrm{Pl}$ with $\mu$: $M_\mathrm{Pl} =
  \mu f_P(\beta_S,\beta_{\sigma},\lambda_S,\lambda_{S\sigma})$,
  where $f_S$ and $f_P$ are dimensionless functions.

\section{Inflation}

\subsection{Effective action for inflation}

 To overcome the problems of old inflation \cite{Guth:1980zm},
at least one bosonic degree of freedom, the inflaton field, must be present 
\cite{Linde:1981mu,Linde:1982zj,Albrecht:1982wi}. 
Despite the fact that the scalaron exists as a bosonic degree of freedom 
in the $R^2$ model inflation \cite{Starobinsky:1980te,Mukhanov:1981xt,Starobinsky:1983zz}, 
it can not generate the spontaneous breaking of scale invariance. 
Similarly, in Higgs inflation \cite{Bezrukov:2007ep},
even though the Higgs field is bosonic, the Coleman-Weinberg mechanism does not work successfully. 
It is for this reason that we have introduced a set of two real scalars, $S$ and $\sigma$, from the start.
The first consequence of spontaneous breakdown of scale invariance
is that the non-minimal coupling to $R$ in ${\cal L}_\mathrm{GR}$
produces  the Einstein-Hilbert term along with
$-(1/2)\beta_S (2 v_S\,S' +S'^2) R$
where $S'=S-v_S$, which suggests that $S$ can play the role of an inflaton
as in the case of Higgs inflation \cite{Bezrukov:2007ep}.
In this section we will pursue this scenario.

Before doing so, we comment on previous literature regarding inflation realized in scale invariant models. While in some Refs.~\cite{GarciaBellido:2011de,Rinaldi:2015uvu, Ferreira:2016wem, Benisty:2018fja, Barnaveli:2018dxo, Ghilencea:2018thl, Kubo:2018kho, Ishida:2019wkd} inflation is not based (explicitly) on a Coleman-Weinberg type potential, other Refs.~\cite{Kannike:2014mia, Kannike:2015apa,Barrie:2016rnv,Vicentini:2019etr, Farzinnia:2015fka, Karam:2018mft, Gialamas:2020snr, Karam:2017rpw} employ the Coleman-Weinberg mechanism to generate the inflaton potential. Most similar to our approach are Refs.~\cite{Farzinnia:2015fka, Karam:2018mft}, even though their inflaton potentials are derived differently. 
Their first step is to go from the Jordan to the Einstein frame, already implicitly assuming that scale invariance is broken, since otherwise the Weyl rescaling is not possible. The resulting Einstein-frame potential consists of two types of scalar fields: Scalar fields stemming from the matter Lagrangian and the scalaron, 
which describes the degree of freedom related to the $R^2$ term i.\e.\ it originates from the gravitational degrees of freedom.
The Gildener-Weinberg approach is then applied to these scalars and quantum corrections are computed in the Einstein frame 
to trigger the breaking of scale invariance.
By contrast, our effective Lagrangian \eqref{Leff}, which includes all 1-loop corrections, is written in the Jordan frame. 
Since the slow-roll parameters are frame independent~\cite{Burns:2016ric,Jarv:2016sow}, it must, in principle, be possible to compute them in the Jordan frame.
Nevertheless, we perform the transformation to the Einstein frame to compute the slow-roll parameters and investigate the slow-roll dynamics explicitly. 
Finally, it should be noted that a transformation between the Jordan and Einstein frames should be taken with care, since as demonstrated in \cite{Kamenshchik:2014waa, Falls:2018olk}, the quantum theories based on the respective classical Lagrangians are not necessarily equivalent. However, this inequivalence only occurs in theories where quantum fluctuations of the metric are included in the 1-loop potential.
Since the present model treats gravity in a purely classical fashion, the metric does not enter the path integral measure and the possible inequivalence is of no concern here.

We now proceed with our case.
As previously noted, we  assume that the higher order terms in Eq.~(\ref{Ueff1}) 
can be  neglected for inflation. We will, however, check throughout whether the aforementioned inequalities $\beta_S  R  < 3 \lambda_S S^2 $ and $\beta_\sigma  R < (1/2) \lambda_{S\sigma} S^2 $
are satisfied.\footnote{
To perform this check, we will use the fact that
the Ricci scalar $R$ during inflation
can be approximated by $12 H^2$, where $H$ is the Hubble
parameter.}
Furthermore, we shall assume that $\kappa$, the coefficient
the Weyl tensor squared term
in Eq.~(\ref{LCGR}), is small enough that it can be neglected for our discussion.
The equation of motion of $S$ does not depend on this term in any case.
With these remarks in mind, we write down
(the relevant part for inflation of) 
the effective Lagrangian  in the Jordan frame,\footnote{%
A similar Lagrangian with \textit{a priori} arbitrary
functions $B$, $G$, and $U$ has been studied in~\cite{Kaneda:2015jma, Canko:2019mud, Gundhi:2018wyz, Gundhi:2020kzm, Gundhi:2020zvb}
for purely phenomenological reasons.
Here we follow Ref.~\cite{Kubo:2018kho}, in which the effective 
Lagrangian is obtained after scale invariance is spontaneously broken
by strong dynamics as proposed in Refs.~\cite{Kubo:2014ova,Kubo:2015cna}.}

\al{
	\frac{{\cal L}_\mathrm{eff}}{\sqrt{-g_J}} 
	=-\frac{1}{2}M_\mathrm{Pl}^2 B(S) R_J+
	G(S) R_J^2+\frac{1}{2}g_J^{\mu\nu}\partial_\mu S\partial_\nu S
	-U_\mathrm{CW}(S)\,,
	\label{Leff}
}
where $g_J^{\mu\nu}\, (g_J=\det g_{\mu\nu}^J)$ and $R_J$ denote the inverse of
the metric $g_{\mu\nu}^J$ and  the Ricci scalar of Jordan-frame spacetime, respectively,
and
\al{ B(S) &=\frac{1}{M_\mathrm{Pl}^2}\left(
	\beta_S S^2+2 U_{(1)}(S)\right)\,\quad\mbox{and}\,\quad
	G(S) =\gamma-U_{(2)}(S)\,.
	\label{BandG}
}
$U_{(1)}(S)$ and $ U_{(2)}(S)$ have been given in Eqs.~(\ref{U1})  and (\ref{U2}),
respectively, and $M_\mathrm{Pl}$ is defined in Eq.~(\ref{mpl}).

We proceed by  introducing an auxiliary field $\psi$ with mass dimension two
to remove the $R_J^2$ term from Eq.~(\ref{Leff}):
\al{
	G(S) R_J^2 &\to 
	2  G(S) R_J\psi-G(S) \psi^2\,.
	\label{aux}}
We then perform a Weyl rescaling of the metric,
$g_{\mu\nu} = \Omega^2\, g_{\mu\nu}^J$ with
\begin{align}
\label{eq:WeylTrafo}
\Omega^2(S,\psi) =
B(S) - \frac{4\,G(S)\psi}{M_{\rm Pl}^2}\,,
\end{align}
to go to the Einstein frame
and arrive at
\begin{align}
\label{eq:LagEO}
\frac{\mathcal{L}_{\rm eff}^E}{\sqrt{- g}} = -\frac{1}{2}\,M_{\rm Pl}^2
\left(R - \frac{3}{2}\,g^{\mu\nu}\,\partial_\mu \ln\Omega^2(S,\psi)\,
\partial_\nu \ln\Omega^2(S,\psi)\right)
+ \frac{g^{\mu\nu}}{2\,\Omega^2(S,\psi)}\,\partial_\mu S\,\partial_\nu S
- V(S,\psi) \,,
\end{align}
where $V$ denotes the scalar potential in the Einstein frame,
\begin{align}
V(S,\psi)  =
\frac{U_\mathrm{CW}(S) + G(S) \psi^2}
{\left[\,B(S) M_{\rm Pl}^2 -
	4\,G(S)\psi\,\right]^2} \, M_{\rm Pl}^4 \,.
\label{VSpsi}
\end{align}
Due to the second term of Eq.~\eqref{eq:LagEO}, $\psi$ is promoted to a propagating scalar field in the Einstein frame.
Its canonically normalized expression, the scalaron field $\phi$~\cite{Barrow:1988xh,Maeda:1988ab}, is defined as
\begin{align}
\label{eq:phi}
\phi = \sqrt{\frac{3}{2}}\,M_{\rm Pl} \ln\left|\Omega^2\right| \,.
\end{align}
The Einstein-frame Lagrangian for the coupled $S$-scalaron system then becomes
\begin{align}
\label{eq:LEphichi}
\frac{\mathcal{L}_{\rm eff}^E}{\sqrt{- g}} = -\frac{1}{2}\,M_{\rm Pl}^2\,R
+ \frac{1}{2}\,g^{\mu\nu}\,\partial_\mu\phi\,\partial_\nu \phi 
+ \frac{1}{2}\,e^{-\Phi(\phi)}\,g^{\mu\nu}\,\partial_\mu
S\,\partial_\nu S
- V(S,\phi)  \,,
\end{align}
where $\Phi\left(\phi\right) = \sqrt{2}\,\phi/\sqrt{3}\,M_{\rm Pl}$, and
the potential $V$ given in Eq.~(\ref{VSpsi}) as a function of $S$ and $\phi$ now reads
\begin{align}
V(S,\phi) = 
e^{-2\,\Phi(\phi)} \left[ U_\mathrm{CW}(S)
+ \frac{M_{\rm Pl}^4}{16\,G(S)}\left(B(S)
- e^{\Phi(\phi)}\right)^2\right] \,.
\label{VSphi}
\end{align}

\subsection{Valley approximation}
\label{sec:valley_approx}
With the scalar potential in Eq.~\eqref{VSphi} at hand, we could proceed by studying the realization of inflation using multifield techniques (see e.g.\ Ref.~\cite{Wands:2007bd}).
We refrain from this complicated approach and instead base our analysis on an effective one-field model to derive predictions for CMB observables. 
This simplification is based on the observation that the scalar potential exhibits a clear valley form along which the potential is relatively flat and thus suitable for slow-roll evolution. Hereafter we will assume that the inflationary trajectory is confined to this valley and that the slow-roll evolution along this contour is parameterized by a single field. This behavior was confirmed in Ref.~\cite{Kannike:2015apa} for a similar model in which the classical trajectories with different initial conditions converge to an inflationary attractor line, i.\e.\ the valley contour.
The existence of a valley form in the scalar potential is
guaranteed as long as a large hierarchy between the two mass eigenvalues of the scalar mass matrix exists. 
This behavior reflects itself as a gradient along the valley which is hierarchically smaller than the gradient perpendicular to it. 
To determine the contour of the valley we use two different approaches and compare them in appendix \ref{app:valley}, where we also show that 
the viability of each approach depends on the region of parameters.

The first approach is based on the observation (see e.g.\ Fig.~\ref{contour_bp2}) 
that there is precisely one local extremum in the scalaron direction for each $S > v_s$ which can be obtained by 
\al{
	\left. \frac{\partial V(S,\phi)}{\partial \phi} \right \rvert_{\phi=\tilde \phi(S)}  = 0 \quad \Rightarrow \quad
	\tilde \phi (S)=\sqrt{\frac{3}{2}}M_\mathrm{Pl}  \ln
	\left( \, B(S)+\frac{16 G(S) U_\mathrm{CW}(S)}{B(S) M_\mathrm{Pl}^4}\,\right)\,,
	\label{phiofS}
}
defining the valley contour in the two-dimensional field space as
\al{\mathcal{C} = \{S,\tilde \phi(S) \}  \quad \mbox{where} \quad \left. \frac{\partial V(S,\phi)}{\partial \phi} \right \rvert_{\phi=\tilde \phi(S)}  = 0 \,.
	\label{contourc}}
Due to the valley structure of the potential $V(S,\phi)$, we may assume
Eq.~(\ref{phiofS}) is approximately satisfied during inflation. The viability of this approximation can be quantified by the requirement
\al{
	\frac{m_\phi^2}{H_\mathrm{inf}^2} \gg 1 \,,
	\label{scalaron_curvature}
}
where $m_\phi$ is the scalaron mass along the contour $\mathcal{C}$ and $H_\mathrm{inf}$ is the Hubble parameter during inflation. If this relation is satisfied, the heavy scalaron mass is able to stabilize the contour $\mathcal{C}$ during the slow-roll phase and the motion in the scalaron direction away from $\mathcal{C}$ can be neglected. 
Inserting $\tilde \phi (S)$ into $ V(S,\tilde \phi(S))$ of Eq.~(\ref{VSphi}) we obtain the inflaton potential along this contour,
\begin{align}
V_{\rm inf}(S) &=  V(S,\tilde \phi(S))=
\frac{U_\mathrm{CW}(S)}{B(S)^2 + 16\,G(S)U_\mathrm{CW}(S)/M_\mathrm{Pl}^4} \,.
\label{Vinf}
\end{align}
Consequently, the kinetic term for $S$ is modified as
\al{
	e^{-\Phi(\tilde{\phi}(S))}\,g^{\mu\nu}\partial_\mu \,S \partial_\nu\, S +
	g^{\mu\nu}\partial_\mu \, \tilde \phi(S) \partial_\nu \,\tilde \phi(S)
	& \Rightarrow F(S)^2 g^{\mu\nu}\partial_\mu \,S \partial_\nu \,S\,, 
	\label{FS}
}
where
\begin{align}
F(S) &= \frac{1}{\left[1+4 \, A(S) \right] B(S)}
\left\{\left[1+4\,A(S)\right]B(S)
+ \frac{3}{2}\,M_{\rm Pl}^2
\left(\left[1+4\,A(S)\right]B'(S)\right.\right.\nn\\
&\left.\left.+ 4\,A'(S)B(S)\right)^2\right\}^{1/2} ~\mbox{with}~
A(S) =\frac{4 G(S) U_\mathrm{CW}(S)}{B(S)^2 M_\mathrm{Pl}^2}\,.
\label{field_norm}
\end{align}
Finally, we arrive at the effective Lagrangian,
\begin{align}
\frac{\mathcal{L}_{\rm eff}^E}{\sqrt{- g}} =
- \frac{1}{2}\,M_{\rm Pl}^2\,R
+ \frac{1}{2}\,F(S)^2\,g^{\mu\nu}
\,\partial_\mu S\,\partial_\nu S
- V_{\rm inf}(S) \,,
\label{LS}
\end{align}
where the canonically normalized inflaton field $\hat{S}$ 
can be obtained from
\begin{align}
\hat{S}(S) = 
\int_{v_S}^S dx \,F(x) \,.
\label{Shat}
\end{align}

The second approach applies if Eq.~\eqref{scalaron_curvature} is violated.
An alternative to obtain the contour then is by looking for local minima in the direction of the field $S$, yielding the contour and inflationary potential,
\al{\mathcal{C}^\prime = \{\tilde S(\phi),\phi \}\,, \quad \mbox{where} \quad \left. \frac{\partial V(S,\phi)}{\partial S}\right \rvert_{S=\tilde S(\phi)} = 0 \,, \quad V_\mathrm{inf}(\phi) = V(\tilde S(\phi),\phi)\,.
	\label{contourcprime}}
Completely analogous to the treatment in the first case, the field normalization (replacing Eq.~\eqref{field_norm}) is 
\al{ F^2(\phi) = \left[1+e^{-\Phi(\phi)} \left( \frac{\partial 
		\tilde{S}(\phi)}{\partial \phi }\right)^2 \right] \,.}

\subsection{One-field description of the slow-roll dynamics}
\label{subsec:numerics}

As shown in the previous section, the two-field system can be
approximately treated as a single field system, either using the contour $\mathcal{C}$ or $\mathcal{C}^\prime$ and the corresponding one-dimensional inflationary potentials (see Eqs. \eqref{Vinf}, \eqref{contourc} and \eqref{contourcprime}). When using the contour $\mathcal{C}$, the canonically
normalized effective inflaton field ${\hat S}$ is defined in Eq.~(\ref{Shat}).
To compute the slow roll parameters it is not necessary to 
use ${\hat S}$ however; instead we employ the following formulae:
\begin{align}
\label{epsilon}
\varepsilon (S) & = 
\frac{M_{\rm Pl}^2}{2\,F^2(S)}
\left(\frac{V'_{\rm inf}
(S)}{V_{\rm inf}(S)}\right)^2 \,,
\\ 
\eta(S) & =
\frac{M_{\rm Pl}^2}{F^2(S)} \left(
\frac{
V''_{\rm inf}(S)}{V_{\rm inf}(S)} -
\frac{ F'(S)}{F(S)}
\frac{V'_{\rm inf}(S)}{V_{\rm inf}(S)}\right) \,.
\label{eta}
\end{align}
The number of e-folds $N_e$ can be computed as
\begin{align}
N_e =\int_{S_*}^{S_\mathrm{end}}
\frac{F^2(S)}{M_{\rm Pl}^2}
\frac{V_{\rm inf}(S)} {V'_{\rm inf}(S)}\,,
\label{efolding}
\end{align}
where ${S_*}$ is the value of
$S$ at the time of CMB horizon exit and $S_\mathrm{end}$  is 
that of $S$ at the end of
inflation, i.\e.\ $\mathrm{max} \{ \varepsilon (S=S_\mathrm{end}), \lvert \eta (S=S_\mathrm{end}) \rvert \} =1$.
The CMB observables, namely the scalar power spectrum amplitude
$A_s$, the scalar spectral index $n_s$ and
the tensor-to-scalar ratio $r$,
can be calculated from
\begin{align}
A_s = \frac{V_{\rm inf\,*}}{24\pi^2\,\varepsilon_*\,M_{\rm Pl}^4} \,, \quad
n_s = 1 + 2\,\eta_* - 6\,\varepsilon_* \,, \quad
r = 16\,\varepsilon_* \,,
\label{parameters}
\end{align}
where the quantities with an asterisk are evaluated at $S=S_*$.
The parameters of our model that are relevant for inflation
are: $\lambda_S, \lambda_{S\sigma},
\beta_S,\beta_\sigma$ and $\gamma$,
where the renormalization scale $\mu$ in the effective potential
(\ref{Ueff}) is fixed through the identification of $M_\mathrm{Pl}$
given in (\ref{mpl}).
We emphasize that all these parameters (except $\mu$) are dimensionless.
Because of the flat direction condition (\ref{flatdr}),
$\lambda_S$ and $\beta_\sigma$ are less relevant and they do not enter $M_\mathrm{Pl}$, see Eq.~(\ref{mpl}), to leading order.
Therefore, we consider the model prediction at a fixed value of $\lambda_S$ and $\beta_\sigma$.
The observables (\ref{parameters}) are measured or constrained by the
latest data from the Planck satellite mission
\cite{Aghanim:2018eyx,Akrami:2018odb}.

For our purpose we assume $N_e \simeq 50 \cdots 60$ e-folds from CMB horizon exit until the end of inflation and constrain the parameter space spanned by
$\lambda_{S\sigma},
\beta_S$ and $\gamma$, such that the following relation is satisfied \cite{Aghanim:2018eyx,Akrami:2018odb}
\al{
\ln (10^{10} A_s) = 3.044 \pm 0.014\;.
\label{As_constraint}
}

\subsection{Numerical analysis of inflation}
\label{sec:pscan}
To discuss the dependence of predictions for CMB observables connected to inflation on the free parameters of the model, we perform a parameter scan using the methods outlined in the previous section. As it turns out, using either the method corresponding to contour $\mathcal{C}$ or $\mathcal{C}^\prime$ has little influence on the prediction for CMB observables. 
Thus, we mainly use contour $\mathcal{C}$, since it allows for an analytic expression of the inflation potential. 
A numeric comparison of the two valley approximations and detailed discussions can be found in appendix \ref{app:valley}.

\begin{figure}[H]
	\begin{center}
		\includegraphics[width=.9\textwidth]{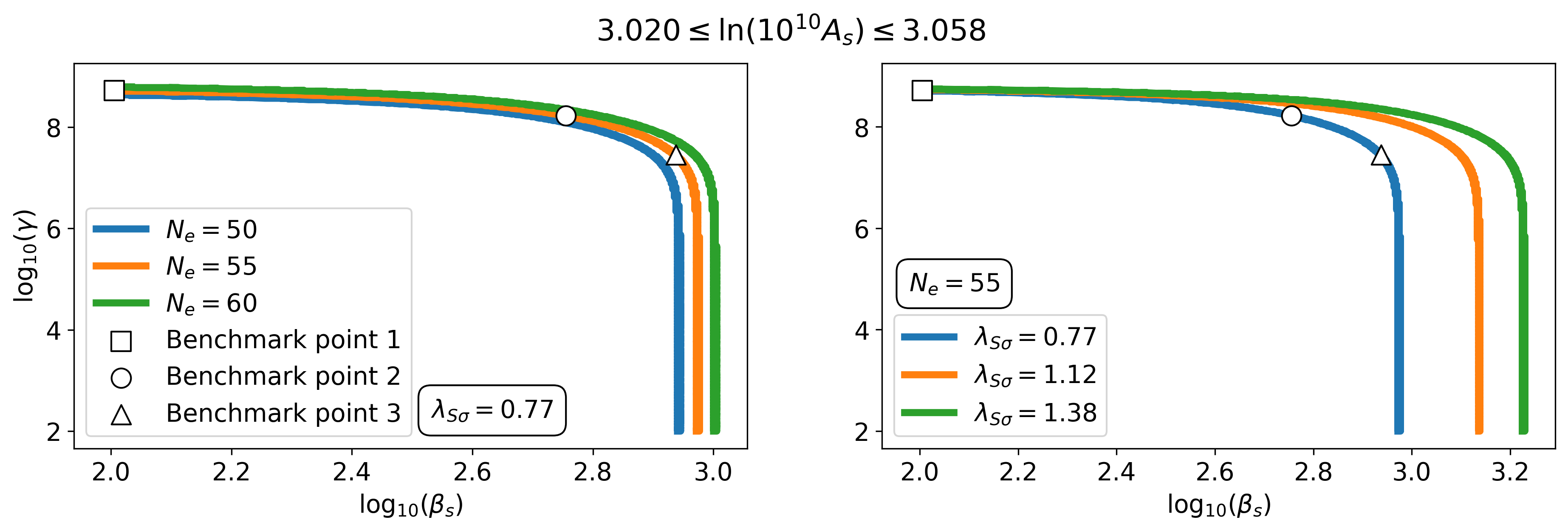}
		\vspace{-2mm}
		\caption{\footnotesize{The lines indicate the parameter combinations of $\gamma$ and $\beta_S$ for which the scalar power spectrum amplitude $A_s$ prediction is fixed to the Planck 
		constraint, Eq.~\eqref{As_constraint}, for a varying number of e-folds $N_e$ (left) or varying $\lambda_{S \sigma}$ (right). 
		For all points we have fixed $\beta_\sigma =1$ and $\lambda_S = 0.005$. The three benchmark points defined in table \ref{tablebp} are marked.}
		}
		\label{As-prediction}
	\end{center}
\end{figure}

\begin{figure}[h]
	\begin{center}
		\includegraphics[width=.7\textwidth]{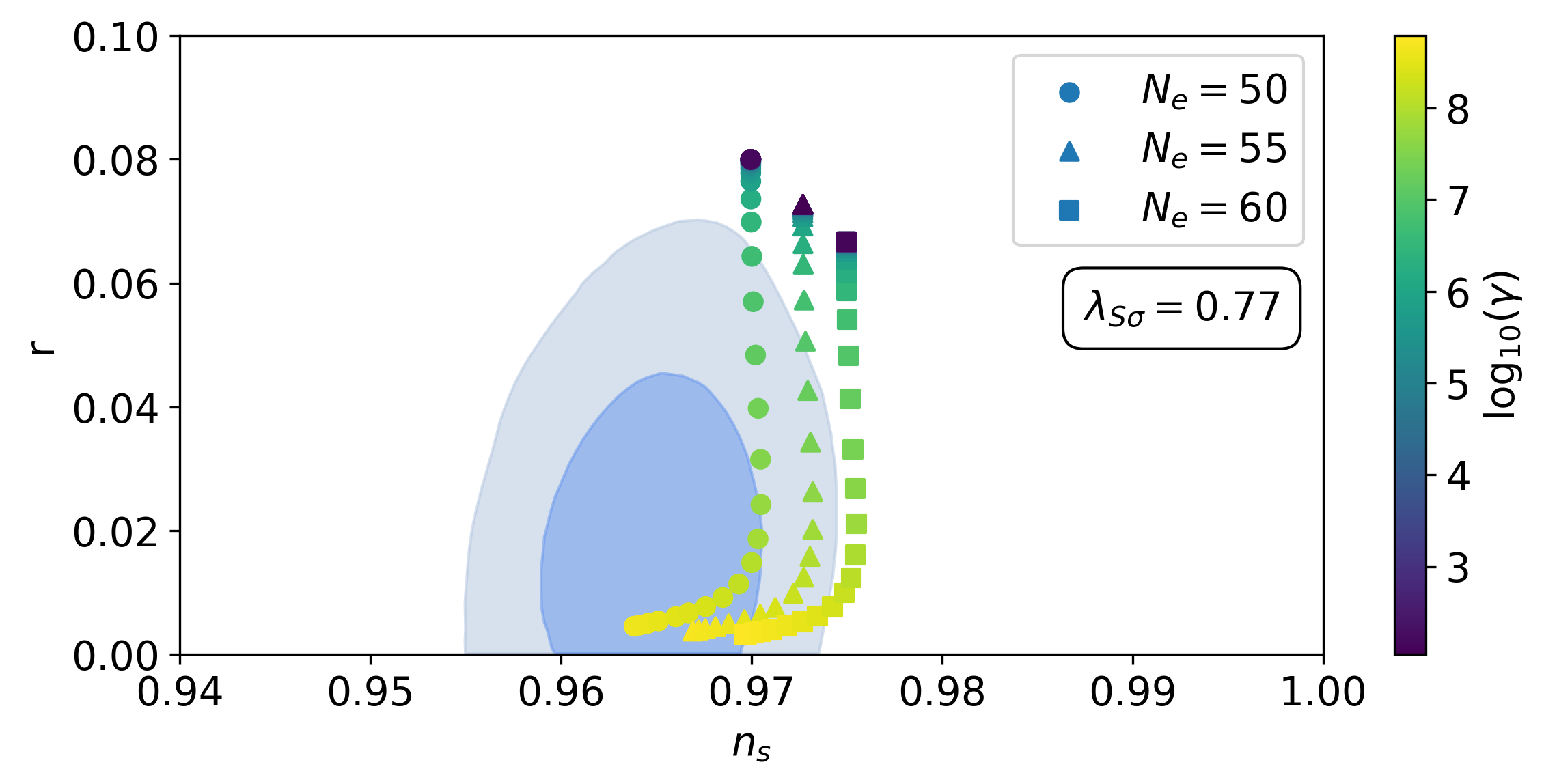} \\
		\includegraphics[width=.7\textwidth]{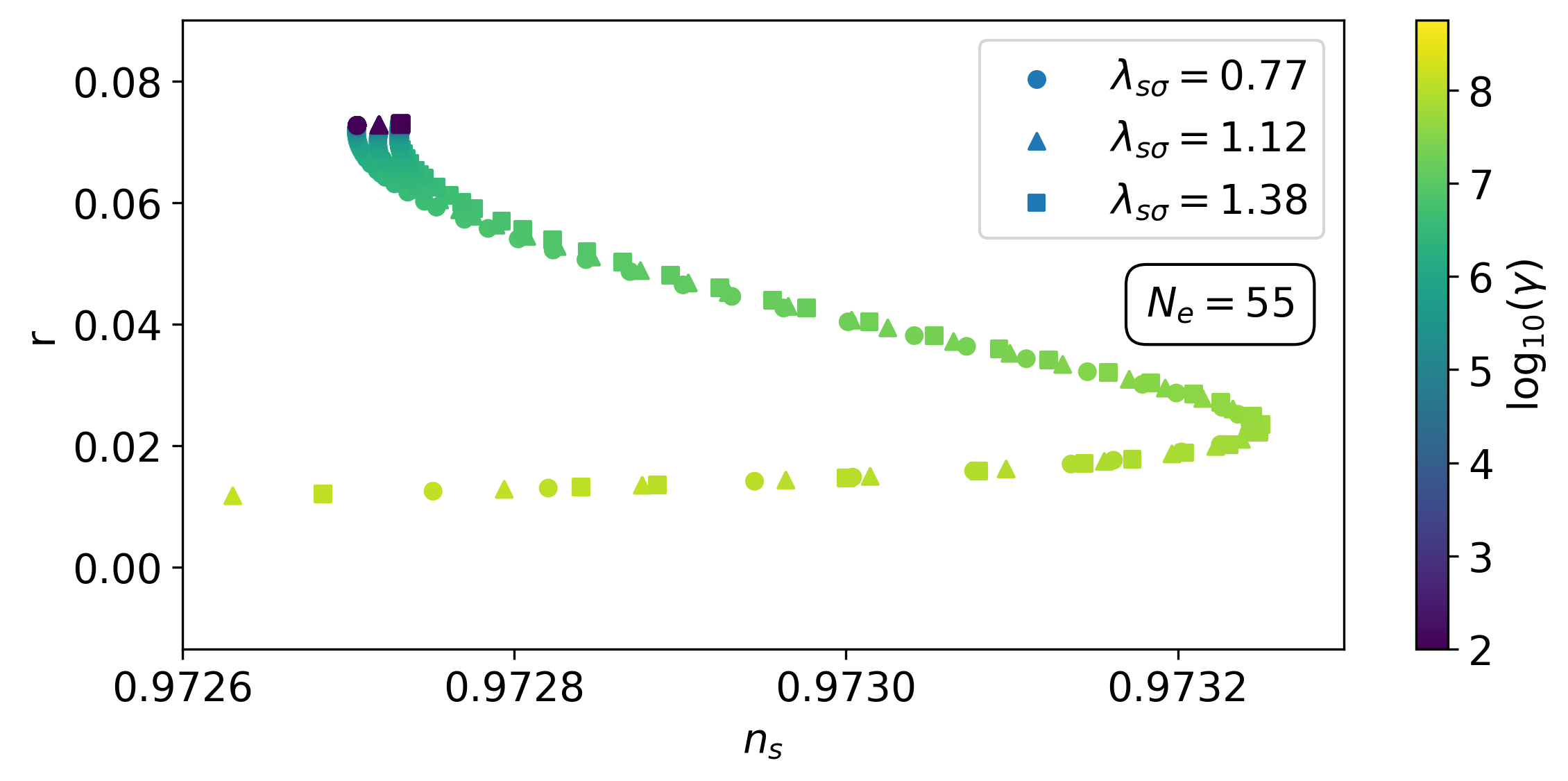}
		\vspace{-2mm}
		\caption{\footnotesize{Predictions for the scalar spectral index $n_s$ and the tensor-to-scalar ratio $r$ with varying number of e-folds $N_e$ (top) and varying $\lambda_{S \sigma}$ (bottom). For all points we have fixed $\beta_\sigma =1$ and $\lambda_S = 0.005$. Only points which satisfy the scalar power spectrum $A_s$ constraint \eqref{As_constraint} are displayed i.\e.\ the $\beta_S$ dependence is fixed in light of Fig.~\ref{As-prediction}. In the top panel we included the Planck TT,TE,EE+lowE+lensing+BK15 $68\%$ and $95\%$ CL regions taken from Ref.~\cite{Akrami:2018odb}.
			}
		}
		\label{ns-r-prediction}
	\end{center}
\end{figure}

For all following results, we have fixed $\lambda_S = 0.005$ and $\beta_\sigma = 1$ as these parameters have little influence on the inflation potential. 
The free parameters left are the portal coupling $\lambda_{S \sigma}$, the $R^2$ coupling $\gamma$ and the non-minimal coupling $\beta_S$. 
Furthermore, the tight observational constraint on the scalar power spectrum amplitude $A_s$ in Eq.~\eqref{As_constraint} can be used to effectively remove one free parameter of the model. 
We use this to obtain a relation between $\beta_S$ and $\gamma$ which is illustrated in Fig.~\ref{As-prediction}. Once $N$ and $\lambda_{S \sigma}$ are fixed, we can express the $\beta_S$ dependence of CMB observables in terms of  $\gamma$ only. One can also see that there are maximally allowed values ($\beta_{S,\mathrm{max}} \sim 10^3$ and $\gamma_\mathrm{max} \sim 10^9$) due to this constraint. The exact values of $\beta_{S,\mathrm{max}}$ and  $\gamma_\mathrm{max}$ depend on $N_e$ and $\lambda_{S \sigma}$.
We utilize this constraint and illustrate the parameter dependence in the $n_s - r$ plane in Fig.~\ref{ns-r-prediction}. All predictions shown there are for points in the parameter space that satisfy 
Eq.~\eqref{As_constraint} (or equivalently, points which are shown in Fig.~\ref{As-prediction}). 
Hence, there is no $\beta_S$ dependence displayed as it is fixed due to the method outlined above. 
As we see from  Fig.~\ref{ns-r-prediction}, the lower end  of the prediction corresponds to that of $R^2$ inflation~\cite{Starobinsky:1980te,Mukhanov:1981xt,Starobinsky:1983zz},
while the upper end is reminiscent of linear chaotic inflation~\cite{Linde:1983gd}\footnote{The results of linear inflation were also reproduced in another context in \cite{Kannike:2015kda, Racioppi:2018zoy}.}
Thus, we see that our predictions interpolate between these two theories.

\begin{table}[H]
\begin{adjustbox}{width=.9\columnwidth,center}
\begin{tabular}{|c|c|c|c|c|c|c|c|c|c|c|c|c|c|c|} 
		\hline
		\multicolumn{3}{|c|}{} & \multicolumn{5}{c|}{Contour $\mathcal{C}$} & \multicolumn{5}{c|}{Contour $\mathcal{C}^\prime$}  \\ 
		\hline
		\#  & $\beta_S$         & $\gamma$  & $n_s$ & $r$ & $A_s$ & $s_\mathrm{end}/\mu$ & $s_*/\mu$ & $n_s$ & $r$ & $A_s$ & $\phi_\mathrm{end}/\mu$ & $\phi_*/\mu$            \\ 
		\hline
		1           & $1.01 \times 10^2 $           & $5.24   \times 10^8$ & 0.967 & 0.004 & 3.032 & 0.09 & 0.11 & 0.965 & 0.004 & 3.088 &  0.83 & 4.75   \\
		2            & $5.69 \times 10^2$           & $1.68 \times 10^8$      & 0.972 & 0.010  &3.041 & 0.11 &  0.45 &0.972 &0.010 & 3.075   & 2.02 & 13.46  \\ 
		3            & $8.67 \times 10^2$            & $2.80 \times 10^7$    & 0.973 &0.034 &  3.038& 0.13 & 2.56 & 0.973&0.034 & 3.040 & 2.74  & 23.46 \\
		\hline
	\end{tabular}
	\end{adjustbox}
\caption{\footnotesize{Parameters of the benchmark points marked in Fig.~\ref{As-prediction}. 
For all points we have fixed $\lambda_{S \sigma} = 0.77$, $\lambda_S = 0.005$ and $\beta_\sigma =1$ i.\e.\ the VEV in each case is $v_S = 0.088 \mu$. 
The last six columns show predictions of CMB observables and related field values by either using the inflaton potential along contour $\mathcal{C}$ or $\mathcal{C}^\prime$ for $N_e = 55$ e-folds. 
See appendix \ref{app:valley} for more details of the valley approximation.}}
\label{tablebp}
\end{table}

\section{Reheating}

During and after the end of inflation the energy density
stored in the inflation is converted to radiation - this process is known as reheating (see, for instance, Refs.~\cite{Kolb:1990vq,Bassett:2005xm}).
Instead of considering a specific model for reheating, we follow~\cite{Liddle:2003as,Martin:2010kz} according to whom 
it is possible to take into account the effect of the reheating phase to some extent without specifying the reheating mechanism. The basic unknown quantities in this approximation are the expansion rate $a_\mathrm{end}/
a_\mathrm{RH}$ of the universe
during the reheating phase and the energy density $\rho_\mathrm{RH}$
at the end of reheating, where $a$ is the scale factor.
These uncertainties can be expressed in a single parameter~\cite{Martin:2010kz}
\al{
R_\mathrm{rad} &=\frac{a_\mathrm{end}}{a_\mathrm{RH}}
\left(\frac{\rho_\mathrm{end}}{\rho_\mathrm{RH}}\right)^{1/4}\,,
\label{Rrad}
}
where $\rho_\mathrm{end}=\rho_S(S_\mathrm{end})$ is the energy density of the inflaton field
at the end of inflation,
and 
$\rho_\mathrm{RH}$ 
is the energy density of radiation at the end of the reheating phase.
The reheating temperature is defined through
\al{
\rho_\mathrm{RH}
&= \frac{\pi^2}{30}\,g_\mathrm{RH} \,T_\mathrm{RH}^4\,,
\label{TRH}
}
where $g_\mathrm{RH}$ corresponds to 
the relativistic degrees of freedom at the end of reheating.
In the following discussion we assume that
$R_\mathrm{rad} $ can be written as \cite{Martin:2010kz}
 \al{
\ln  R_\mathrm{rad} &=\frac{1-3 \bar{w}}{12(1+\bar{w})}\ln \left(\frac{\rho_\mathrm{RH}}{\rho_\mathrm{end}}\right)\,,
\label{Rrad1}
 }
where $\bar{w}$ is the average equation of state parameter in the reheating phase.

Next, we constrain the number of e-folds
$N_e=\ln \left(a_\mathrm{end}/a_*\right)$.
Here $a_*$ is the scale factor at the time of CMB horizon exit and
is defined as $k_*=a_* H_*$, where $k_*$ is the pivot scale set by the Planck
mission \cite{Aghanim:2018eyx,Akrami:2018odb}, 
and $H_*$ is the Hubble parameter at $a=a_*$.
One finds~\cite{Martin:2010kz,Martin:2013tda}\footnote{%
Eq.~(\ref{Ne})  can be derived from 
$a_\mathrm{end}/a_*=R_\mathrm{rad} \,
\left(  a_\mathrm{RH} \rho_\mathrm{RH}^{1/4} /
\sqrt{3} a_0 H_0\right) \left(\sqrt{3} H_*/\rho_\mathrm{end}^{1/4}\right)
\left(a_0 H_0/k_*\right)$, where $R_\mathrm{rad}$ is defined in
Eq.~(\ref{Rrad}).
}
\al{
N_e &=\ln \left(\frac{a_\mathrm{end}}{a_*}\right)=\ln \left(\frac{a_\mathrm{RH}\,\rho_\mathrm{RH}^{1/4}}{\sqrt{3} a_0 \,H_0} \right)-\ln \left(\frac{k_*}{a_0 H_0}  \right)+\frac{1}{4}
\ln \left(\frac{V_\mathrm{inf\,*}^2}{M_\mathrm{Pl}^4 \,\rho_\mathrm{end}}  
\right) + \ln \left( R_\mathrm{rad} \right)\,,
\label{Ne}
}
where $a_0=1$ and $H_0=h \,
2.13\times 10^{-42}$ GeV  are the present values of the scale factor
and the Hubble parameter, respectively, $k_*\,=\,0.002\, \mbox{Mpc} ^{-1} $ \cite{Aghanim:2018eyx},
and $g_\mathrm{RH}=106.75+(7/8) 12=117.25$. The first term of Eq.~(\ref{Ne}) 
can be computed by using Eq.~(\ref{TRH}) and
the conservation of entropy
$a_\mathrm{RH}^3 \,s_\mathrm{RH} =a_0^3\, s_0$,
giving $a_\mathrm{RH}/a_0 = (q_0/q_\mathrm{RH})^{1/3}\, T_0/T_\mathrm{RH}$,
where $q_0=43/11$, and $q_\mathrm{RH}=g_\mathrm{RH}$
are the degrees of freedom that enter via entropy
at the present day and at the end of the reheating phase, respectively.
Then, using $T_0=2.725\,K$ one finds 
\cite{Lozanov:2017hjm,Akrami:2018odb}
\al{
\ln \left(\frac{a_\mathrm{RH}\,\rho_\mathrm{RH}^{1/4}}{\sqrt{3} a_0 \,H_0} \right)
&=
66.89-\frac{1}{12}\ln g_\mathrm{RH}\,.
\label{67}
}
The energy density at the end of inflation $\rho_\mathrm{end}$ can be expressed
in terms of the slow-roll parameter as
\al{
\rho_\mathrm{end} &=
\frac{V_\mathrm{end}(3-\varepsilon_*)}{(3-\varepsilon_\mathrm{end})}\,,
}
where $V_\mathrm{end}=V(S_\mathrm{end},\phi(S_\mathrm{end}))$,
$\varepsilon_\mathrm{end}=\varepsilon(S_\mathrm{end})$, and
$\varepsilon_*=\varepsilon(S_*)$.
The average equation of state parameter $\bar{w}$ in Eq.~(\ref{Rrad1})
can be found from the behavior of the scalar potential $V(S,\phi)$
near the potential minimum. Noticing from Eq.~(\ref{BandG}) together with Eqs. (\ref{U1}) and (\ref{U2}) that
$B(S)\simeq 1+O(S-v_S)\,,\, G(S) \simeq \gamma+\dots$ 
and also that $\phi(S)\simeq O(S-v_S)$, we have
$\exp\left[\,\Phi(\phi(S))\,\right]\simeq 1+O(S-v_S)$  near $S=v_S$, and so
we  find 
\al{
V(S,\phi(S)) &\simeq U_\mathrm{CW}(S)
= (3 \lambda_S v_S^2+\dots)\hat{S}^2+O(\hat{S}^3)
\label{V1}
}
where $\hat{S}$ 
$\simeq S-v_S$
(because $F(S) \simeq 1+O(S-v_S)$ which can be understood 
from Eq.~(\ref{FS}) near $S=v_S$). Here,
``$\dots$'' stands for the higher order contribution
in the effective potential Eq. (\ref{Ucw}).
The constant term  $V(v_S,\phi(v_S))$ is absent because we have 
subtracted the zero-point energy $U_0$. The term linear to $\hat{S}$
is also absent because $\hat{S}=0$ is the position of the minimum
of $U_\mathrm{CW}$. Therefore, we deduce from Eq.~(\ref{V1}) 
that $p=2$
and $\bar{w}=(p-2)/(p+2)=0$ \cite{Turner:1983he}.\footnote{%
$p$ is defined from the behavior of the potential
near the minimum: $V\sim\hat{S}^p$.}
Therefore, once the slow roll parameters and the pivot scale $k_*$
are fixed, 
the only quantity on rhs of Eq.~(\ref{Ne}) 
that is not free is the reheating temperature $T_\mathrm{RH}$.
That is, Eq.~(\ref{Ne}) can be understood as a constraint on $N_e$, assuming that
$(1\,\mbox{TeV})^4 \lsim \rho_\mathrm{RH} \lsim \rho_\mathrm{end}$ is satisfied~\cite{Akrami:2018odb}.
On the other hand, since $49 < N_e < 59$ must also be satisfied~\cite{Akrami:2018odb},
Eq.~(\ref{Ne}) gives a constraint on $T_\mathrm{RH}$
if $\rho_\mathrm{RH}\in [1\,(\mbox{TeV})^4,\rho_\mathrm{end}]$ 
is simultaneously satisfied.
As we see later when discussing dark matter (and also briefly leptogenesis), 
the reheating temperature $T_\mathrm{RH}$ plays an important role.

\begin{figure}[t]
	\centering
	\includegraphics[width=.75\textwidth]{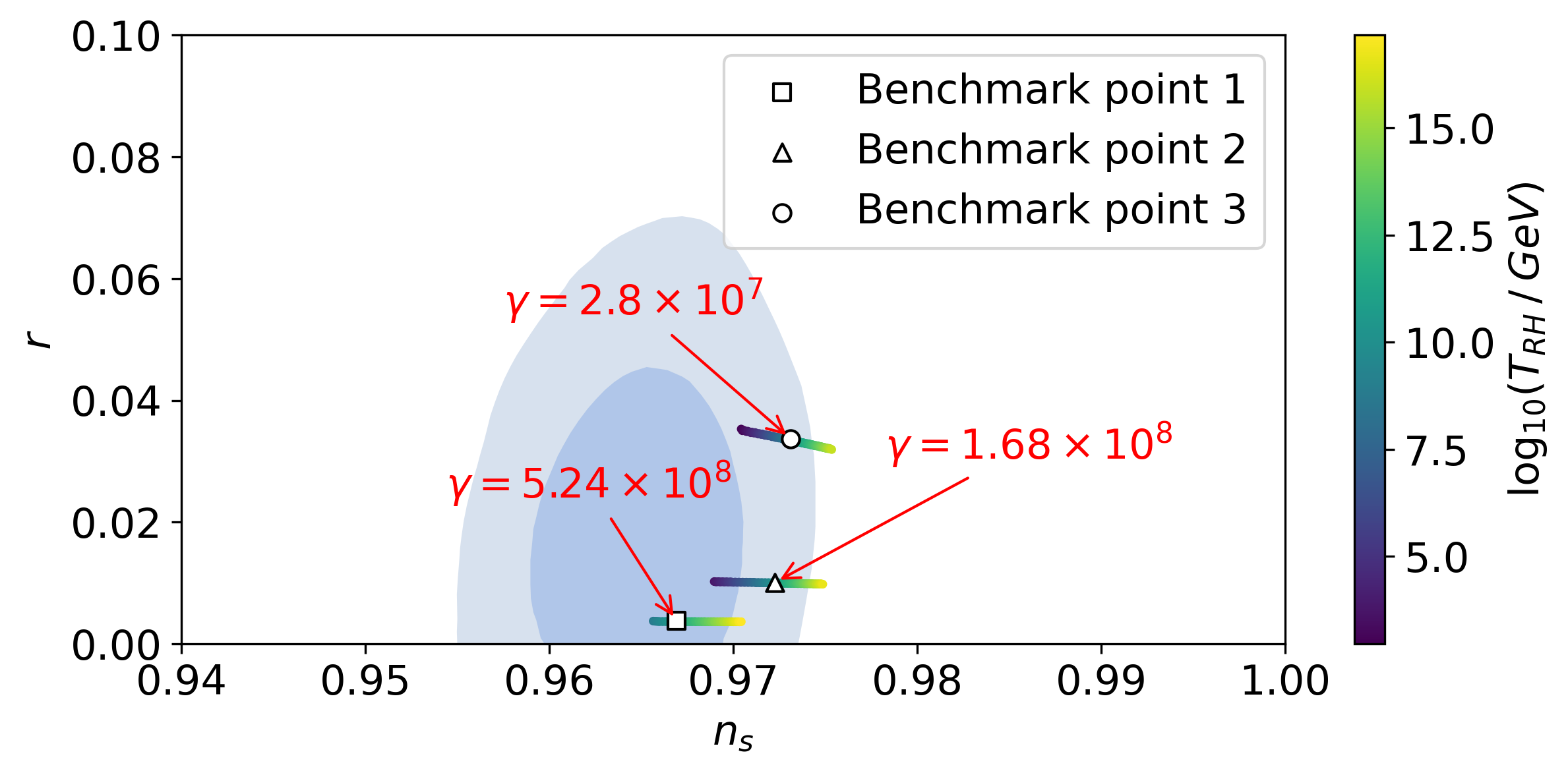}
	\vspace{-2mm}
	\caption{\footnotesize{Predictions for the scalar spectral index $n_s$ and the tensor-to-scalar ratio $r$ with varying number of e-folds $N_e \in [50,60]$ and (slightly) varying $\beta_S$ to account for the constraint on $A_s$ from Eq.~(\ref{As_constraint}). $T_{\mathrm{RH}}$ is shown using its dependency on the number of e-folds $N_e$, from Eq.~(\ref{efolding}). For all points we have fixed $\lambda_S = 0.005, \; \lambda_{S \sigma} = 0.77, \; \beta_\sigma =1 $ and $\gamma$ for each line respectively as seen in the figure. 
	We also show the Planck TT,TE,EE+lowE+lensing+BK15 $68\%$ and $95\%$ CL regions taken from Ref.~\cite{Akrami:2018odb}.
			}}
	\label{fig:InfParametersTreheating}
\end{figure}

We have used the relation between $N_e$ and $T_{\mathrm{RH}}$ in Eq.~(\ref{Ne}) to demonstrate how varying $N_e$ effects the inflation parameters ($r, \: n_s$) via the corresponding reheating temperatures in Fig.~\ref{fig:InfParametersTreheating}. Here all couplings are fixed to benchmark point $1,2$ or $3$ (see table~\ref{tablebp}), but we vary the assumed $N_e \in [50,60]$ and adjust $\beta_S$ (slightly) such that the constraint on $A_s$ is fulfilled. Because of the lower bound on $\beta_S$, we only display $N_e \in \left[ 53.5, 60\right] $ in the line corresponding to benchmark point 1. The reheating temperature $T_{\mathrm{RH}}$ is then shown via color-scaling on the usual $n_s-r$ plot (see Fig.~\ref{As-prediction}). Note that any resulting constraints on $T_{\mathrm{RH}}$ can be converted into constraints on $N_e$, and vice versa, using the aforementioned relation from Eq.~(\ref{Ne}).

\section{Dark matter}
\begin{figure}[t]
\begin{center}
\includegraphics[width=2.4in]{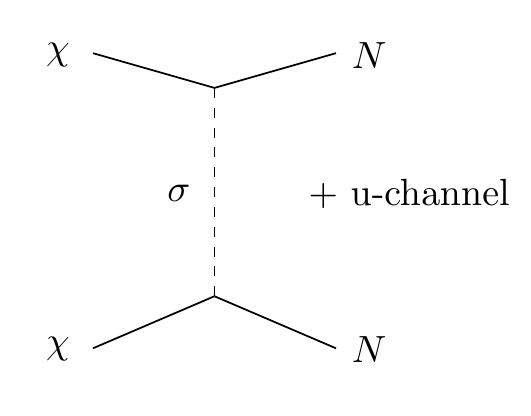}
\includegraphics[width=2.7in]{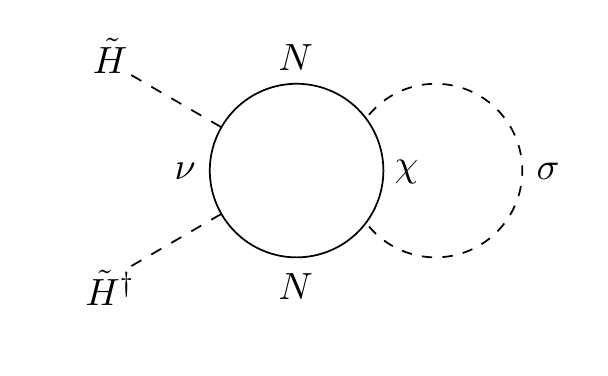}
\vspace{-2mm}
\caption{\footnotesize{Feynman diagramms for the scattering process $\chi\chi \leftrightarrow NN$ (left) and two-loop contribution to the Higgs mass term (right).
}}
\label{XXNN}
\end{center}
\end{figure}
Since the discrete $Z_2$ symmetry is not spontaneously broken, the $Z_2$-odd particles $\sigma$ and $\chi$ are stable and, therefore, 
good dark matter candidates. 
Dark mater can be produced during or after the reheating phase, see e.g.~\cite{Chung:1998rq,Allahverdi:2002nb,Garcia:2020eof}.
Because of the inequality Eq.~(\ref{flatdr}), which is needed to realize the desired
flat direction, $\sigma$ is always heavier than $S$.  
This implies that the inflaton $S$ cannot decay into $\sigma$, and $\sigma$ is not produced during the reheating stage~\cite{Allahverdi:2002nb}.
In contrast to $\sigma$, $\chi$ can be produced; either through the inflaton decay $S \to \chi\,\chi$ or
by the scattering process $N\,N \to \chi\,\chi$.
The corresponding diagram is shown in the left panel of 
Fig.~\ref{XXNN},\footnote{%
There is also an s-channel diagram for the process 
$N\,N \to \chi\,\chi$, but this process 
is absent if the scalar field $S$ is treated as a classical field.
Even if the process exists, the decay $S\to\chi\,\chi$
is dominant due to the fact that the former process is proportional to
$y_\chi^2 y_M^2$ while the later one to $\propto y_\chi^2$.}
where a thermal abundance of right-handed neutrinos $N$ is assumed to exist since $N$ has contact 
with the SM sector through the Dirac-Yukawa coupling $y_\nu$.
However, the cross section $\sigma_{N\chi}
\sim (y_{N\chi}^4/\pi)  \,\mathrm{max} 
\{ m_N^2,m_\chi^2\}
/m_\sigma^4$, which corresponds to the the scattering process in Fig.~\ref{XXNN},
is extremely suppressed simply due to the fact that $\sigma$ is very heavy i.\e.\
$m_\sigma \sim 10^{-2} M_\mathrm{Pl}$, while
 $m_\chi=v_S\, y_\chi$ and $v_S= m_N /y_M$. This leads to
\al{\label{chichiNN}
\sigma_{N\chi} \sim \begin{cases}
	10^8\,\left(\frac{y_{N\chi}^4\,y_\chi^2}{\pi \,y_M^2}\right)\left(\frac{m_N^2}{M_\mathrm{Pl}^4}\right) \sim \left(\frac{y_{N\chi}^4\,y_\chi^2}{\pi}\right)\,\left[10^{16}\mbox{GeV}\right]^{-2} \,, & \mathrm{for}\quad m_N < m_\chi\,, \\
	10^8\,\left(\frac{y_{N\chi}^4}{\pi }\right)\left(\frac{m_N^2}{M_\mathrm{Pl}^4}\right) \sim \left(\frac{y_{N\chi}^4}{\pi}\right)\,\left[10^{25}\mbox{GeV}\right]^{-2} \,, & \mathrm{for}\quad m_N >  m_\chi\,,
\end{cases}
}
where we have used $m_N=10^7$ GeV, 
$M_\mathrm{Pl}=2.43\times 10^{18}$ GeV and
$y_M=m_N/v_S\simeq  10^{-10}$.
Furthermore, $y_{N\chi}$ can be constrained due to the two-loop diagram
(shown in the right panel of Fig.~\ref{XXNN})
that contributes to the Higgs mass term $\Delta \mu_H$,
which should be much smaller than $O(10^2)$ GeV in order to realize the neutrino option.
A rough estimate of the two-loop diagram
gives $\delta\mu_H\sim y_\nu \,y_{N\chi}
\,m_\sigma/16\pi^2$, from which we find
$y_{N\chi}\ll  O(10^{-8})$. Inserting this into 
(\ref{chichiNN}), we immediately find  that $\sigma_{N\chi}$
is too small to be relevant for the production of 
$\chi$ before, as well as after the end of reheating \cite{Chung:1998rq}.
Therefore, we ignore this process in the following discussion
and concentrate on the decay of $S$ into two $\chi$. 
The corresponding decay width is given by
\al{
\gamma_\chi &=
\frac{3 \,y_\chi^2m_S}{16\pi} (1-4 m_\chi^2/m_S^2)^{1/2}\,.
\label{gammachi}
}
Note that $\chi$ has  contact with the SM particles only through 
$N$.  Therefore, because of the constraint
$y_{N\chi} \ll  O(10^{-8})$, its contact with the SM is extremely suppressed.

\begin{figure}[t]
	\begin{center}
		\includegraphics[width=\textwidth]{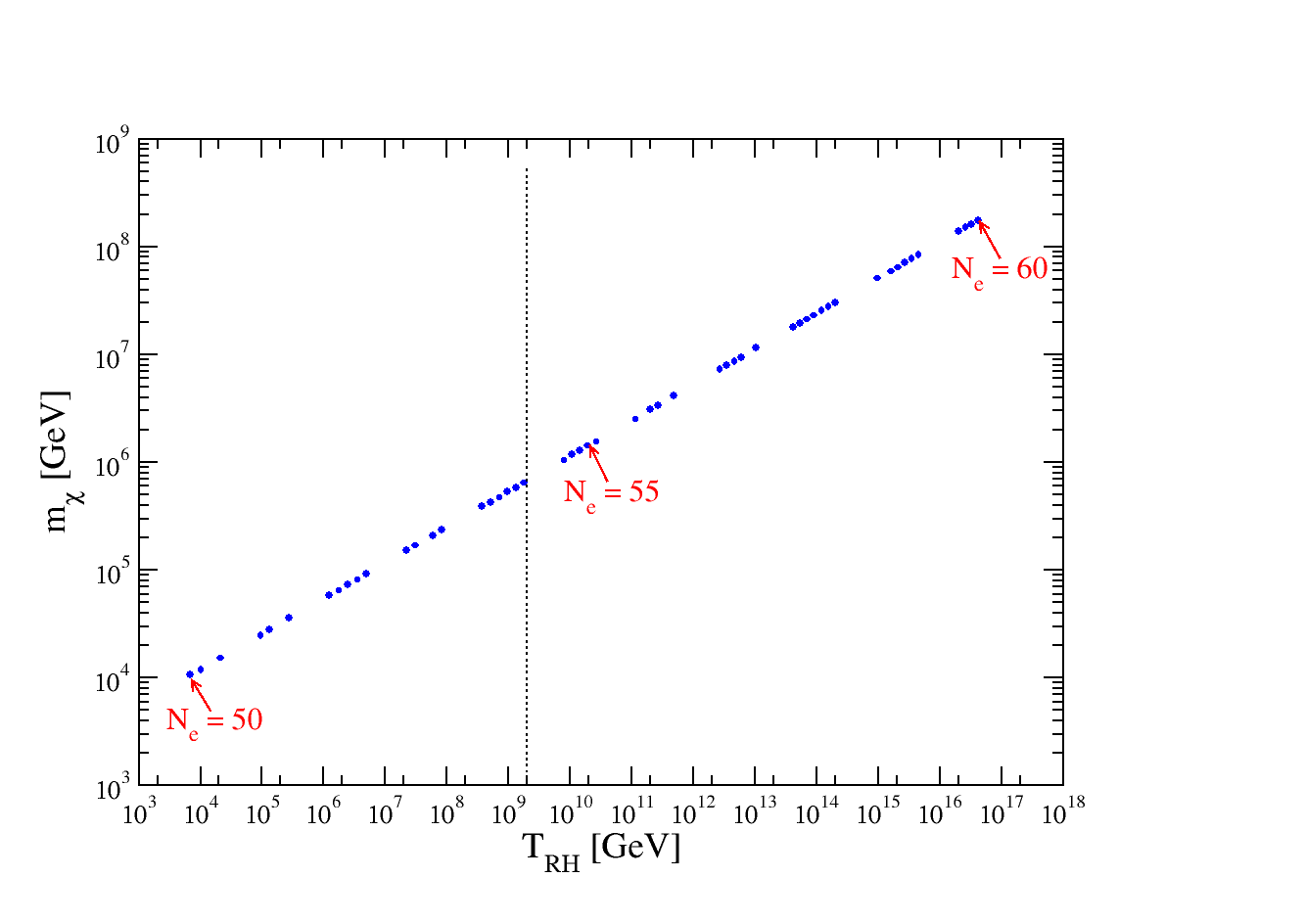}
		\vspace{-2mm}
		\caption{\footnotesize{Dark matter mass $m_\chi$ against reheating temperature $T_\mathrm{RH}$.
	We have varied  $\beta_S$ around the value $5.69\times 10^2$ of benchmark point 2 (see table~\ref{tablebp}), while all the other input parameters are fixed to 
	the benchmark point values. The result is found to be quite insensitive against the change of these parameters. The black dotted line shows the lower bound on $T_\mathrm{RH}$ 
	for a viable thermal leptogenesis with $m_N \gtrsim 2\times10^7$ GeV~\cite{Giudice:2003jh}.
			}
		}
		\label{TRH-DM}
	\end{center}
\end{figure}

With the assumptions above we finally arrive at a system\footnote{%
Here we assume that $S$ is the dominant part 
of the inflaton field, which is a mixture of $S$ and $\phi$ in general.
If the mixing is large, one can incorporate it into
the decay width (\ref{gammachi}).} 
containing only the inflaton $S$ and the dark matter field $\chi$. 
We can now consider the coupled Boltzmann equations~\cite{Chung:1998rq}
\al{
\frac{d n_S}{dt} &= -3 H n_S-\Gamma_S\, n_S \,,
\label{nS}\\
\frac{d n_\chi }{dt}&= -3 H n_\chi+ B_\chi \Gamma_S \,n_S \,,
\label{nchi}
}
where $n_\chi$ stands for the  number density of
$\chi$,   the energy density of $S$ is denoted by $\rho_S=m_S \,n_S$, $B_\chi=\gamma_\chi/\Gamma_S$, and
$\Gamma_S$ is the total decay width of $S$.
Eq.~(\ref{nS}) is not coupled and can be solved \cite{Kolb:1990vq} to find
\al{
n_S (a) &= \frac{\rho_\mathrm{end}}{m_S}\,\left[
\frac{a_\mathrm{end}}{a}\right]^3\, e^{-\Gamma_S\,(t-t_\mathrm{end})}\,,
\label{nSa}
}
where $a$ is the scale factor at $t > t_\mathrm{end}$, $a_\mathrm{end}$
is $a$ at the end of inflation $t_\mathrm{end}$ and 
$\rho_\mathrm{end}=\rho_S(a_\mathrm{end})$.
To solve Eq.~(\ref{nchi}) we insert the solution 
(\ref{nSa}) and find
\al{
n_\chi (a) &=
B_\chi \,\frac{\rho_\mathrm{end}}{m_S}\,
\left[\frac{a_\mathrm{end}}{a}\right]^3\,\left(1- e^{-\Gamma_S\,(t-t_\mathrm{end})}\right)\,.
}
The freeze-in value of $n_\chi$ is the value at $t=\infty$, which implies that
 the relic abundance $\Omega_\chi h^2$ is given by
\al{
\Omega_\chi h^2 &=
m_\chi\,B_\chi \,\frac{\rho_\mathrm{end}}{m_S}\,
\left[\frac{a_\mathrm{end}}{a_0}\right]^3\, 
\frac{ M_\mathrm{Pl}^2}{3 (H_0/h)^2}\,,
}
where  $a_0=1$  and $H_0= h\,2.1332\times 10^{-42}$ GeV
with $h\simeq 0.674$ \cite{Aghanim:2018eyx}
stand for the present value of the scale factor and the
Hubble parameter, respectively.
Note that $a_\mathrm{end}/a_0$  can be computed similarly to the derivation of Eq.~(\ref{Ne}),
\al{
\frac{a_\mathrm{end}}{a_0} &=
\left(\frac{a_\mathrm{end}}{a_\mathrm{RH}}\right) ~
\left(\frac{\rho_\mathrm{end}^{1/4}}{\rho_\mathrm{RH}^{1/4}}\right) ~
\left(\frac{a_\mathrm{RH}\,\rho_\mathrm{RH}^{1/4}}{\sqrt{3} a_0 \,H_0} \right)
\left(\frac{\sqrt{3} \,H_0}{\rho_\mathrm{end}^{1/4}}  \right)\,.
\label{aend}
}
As we see from Eq.~(\ref{aend}), the product of the first two expressions
is $R_\mathrm{rad}$ defined in Eq.~(\ref{Rrad}) and
 the  quantity in the third parenthesis is 
exactly $\exp( 66.89-\ln g_\mathrm{RH}/12)$.
Note that the $\rho_\mathrm{end}$ dependence in $\Omega_\chi
h^2$ cancels because the average equation of state $\bar{w}$
is zero in our case (see the discussion
below Eq.~(\ref{V1})), and also $g_\mathrm{RH}$ cancels if Eq.~(\ref{TRH}) is used for $\rho_\mathrm{RH}$.
We then find
\al{
\Omega_\chi h^2 &=
\sqrt{3} \exp(3\times 66.89)\,\frac{B_\chi \,H_0}{M_\mathrm{Pl}^2}\,
\left(\frac{\pi^2}{30}\right)^{1/4}\,
\left(\frac{m_\chi}{m_S}\right)\,T_\mathrm{RH}\,\\
&\simeq 2.04\times 10^{8}\,B_\chi \left(\frac{m_\chi}{m_S}\right) \,\frac{T_\mathrm{RH}}{1\,\mbox{GeV}}\,,
}
 which is in accordance with the results of Ref.~\cite{Allahverdi:2002nb}.
The branching ratio $B_\chi=\gamma_\chi/\Gamma_S$ can be obtained
by using Eq.~(\ref{gammachi}) for $\gamma_\chi$ and by assuming that
$1/\Gamma_S$ can be identified with the time scale at the end of the reheating phase~\cite{Kolb:1990vq,Chung:1998rq} i.\e.\ $1/H(a_\mathrm{RH})
=\left(\,3\, M_\mathrm{Pl}^2/\rho_\mathrm{RH}
\,\right)^{1/2}$. 
For the benchmark point 2 in table \ref{tablebp} ($m_S\simeq4.4\times 10^{15}\,\mbox{GeV}\,,v_S\simeq 1.0\times 10^{17}\,\mbox{GeV}\,,
T_\mathrm{RH}\simeq 1.9\times 10^{10}\,\mbox{GeV}\,,
k_*=0.002\, \mbox{Mpc} ^{-1}$) we obtain
\al{
\Omega_\chi h^2&\simeq 4.4\times 10^{31}~
y_\chi^3 \simeq 0.12\,,\quad\mbox{for}\quad
y_\chi\simeq 1.4\times 10^{-11}\,,
}
from which we also find that
$m_\chi=y_\chi\, v_S\simeq 4.3\times 10^{6}$ GeV.
In Fig.~\ref{TRH-DM} we plot $m_\chi$ against $T_\mathrm{RH}$, where we have varied $\beta_S$ from the benchmark point
value $ 5.69\times 10^2 $
(with all the other input parameters fixed to the benchmark point values).
The interval of  $\beta_S$ is chosen such that $N_e$ varies between $50$ and $60$, where 
$T_\mathrm{RH}=6.8\times 10^3$ GeV at $N_e=50$ and $T_\mathrm{RH}=4.1\times10^{16}$ GeV at $N_e=60$. 
The reheating temperature $T_\mathrm{RH}$ changes considerably as $N_e$ changes. 
This can be understood using the fact that $N_e = (1/3)\ln T_\mathrm{RH}+\dots$, as one can see from Eq.~(\ref{Ne}) together 
with Eqs.~(\ref{TRH}) and (\ref{Rrad1}).
Accordingly, $m_\chi$ (and also  $y_\chi$) varies quite a lot. 
The black dotted line is the lower bound on 
$T_\mathrm{RH}$  for a viable thermal leptogenesis with $m_N \gsim 10^7$~GeV~\cite{Giudice:2003jh}.

\section{Neutrino option}
\begin{figure}[t]
	\begin{center}
		\includegraphics[width=2.5in]{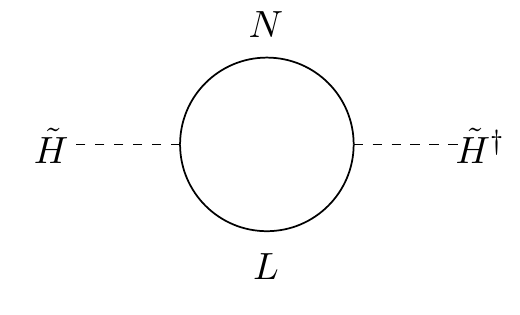}
		\includegraphics[width=2.2in]{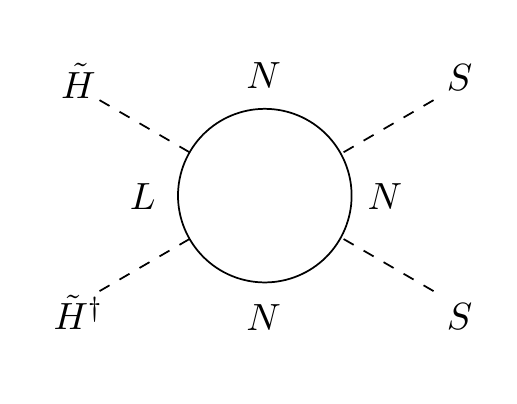}
		\vspace{-2mm}
		\caption{\footnotesize{Neutrino contributions
				to the Higgs mass term (left) and Higgs portal coupling (right).
			}
		}
		\label{HHSS}
	\end{center}
\end{figure}
Heavy right-handed neutrinos introduce important corrections to the Higgs mass term, $-\mu_H^2 H^\dag H$,
due to the diagram shown in Fig.~\ref{HHSS}~(left). 
The finite term  of this contribution, which, in general, depends on the renormalization scale, 
is given by~\cite{Vissani:1997ys,Casas:1999cd,Clarke:2015gwa,Bambhaniya:2016rbb}
\begin{align}
|\Delta \mu_H^2| \sim \frac{y_\nu^2 m_N^2}{4\pi^2}\,.
\label{eq:higgs}
\end{align}
The neutrino option, as presented in
\cite{Brivio:2017dfq},  assumes that 
(\ref{eq:higgs}) is the dominant contribution
for the Higgs mass term, i.\e.\ $\Delta \mu_H^2 \sim\mu_H^2
\simeq 2 (125\, \mbox{GeV})^2$.
Requiring at the same time that the light SM neutrino masses are generated via a type-I seesaw mechanism~\cite{Minkowski,GellMann:1980vs,Yanagida:1979as,Goran}, i.\e.\
$m_\nu \simeq y_\nu^2 v_h^2/m_N \sim 0.1$ eV
with $v_h=246$\,GeV, we find that $m_N$ is of order $10^7$ GeV and $y_\nu$  is of order $10^{-4}$~\cite{Brivio:2017dfq}(see also Refs.~\cite{Brivio:2018rzm,Davoudiasl:2014pya}). 
In Refs.~\cite{Brdar:2018vjq,Brdar:2018num,Aoki:2020mlo} the original model of Ref.~\cite{Brivio:2017dfq} has been
embedded into a classically scale invariant theory.
In the scale invariant extension,
$\Delta \mu_H^2=0$ before spontaneous scale symmetry breaking while radiative correction to the dimensionless coupling $\lambda_{HS}$ exist, 
mainly due to the diagram shown in Fig.~\ref{HHSS}~(right). 
Consequently, for the neutrino option to work, we must assume that the one-loop correction proportional to $y_\nu^2 y_M^2/16\pi^2$ 
is the main origin of $\lambda_{HS}$. After the spontaneous breaking of scale symmetry we then obtain $\Delta \mu_H^2\sim  y_\nu^2 y_M^2 v_S^2/4\pi^2$.
In the present model of a unified origin of energy scales, the origin of $m_N$
is the therefore the same as for $M_\mathrm{Pl}$:
\al{
	m_N &= y_M v_S =
	y_M M_\mathrm{Pl}\left(\beta_S+2 U_{(1)}(v_S)/v_S^2\right)^{-1/2}\,,
}
where the analytic expression for $2 U_{(1)}(v_S)$ is given in Eq.~(\ref{mpl}).
Since $\beta_S\gg 2 U_{(1)}(v_S)/v_S^2$ is satisfied in the parameter space we consider, we find
\al{
	y_M \simeq \frac{m_N \beta_S^{1/2}}{M_\mathrm{Pl}}
	\simeq 10^{-10 }
	\times\left(\frac{\beta_S}{10^3}\right)^{1/2}\,.
	\label{yM}
}
As already mentioned in the introduction,
the smallness of $y_M$ is technically natural in the sense of 't Hooft \cite{tHooft:1979rat}, because 
$U(1)_{B-L}$ is  restored as $y_M$ and $y_{N\chi}$ go to zero (recall that in the previous section when discussing dark matter, we argued
that $y_{N\chi}\ll 10^{-8}$).

At last we would like to emphasize that leptogenesis~\cite{Fukugita:1986hr,Buchmuller:2004nz}, 
i.\e.\ the generation of the baryon asymmetry of the Universe works successfully within the frame work of the neutrino option~\cite{Brdar:2019iem,Brivio:2019hrj}. 
If we assume that the right-handed neutrinos $N$ can be reheated only through the contact with the SM particles,\footnote{%
The direct reheating is very small because the coupling $y_M$ of $N$ to $S$ is very small as seen from Eq.~(\ref{yM}).}
the bound $T_\mathrm{RH}\gsim 2\times 10^9$  GeV must be satisfied in order to realize
thermal leptogenesis with  $m_N \gsim 2\times 10^7$ GeV \cite{Giudice:2003jh}.
This lower bound on 
$T_\mathrm{RH}$  is shown in Fig.~\ref{TRH-DM} (black dotted line).
For the three benchmark points in table~\ref{tablebp}, we find that thermal leptogenesis works under our assumptions 
only if $N_e \gsim 54$.

\section{Conclusions}
We have investigated a framework for unifying the origin of the fundamental energy scales in Nature, namely the Planck and electroweak scales, utilizing a classically scale invariant model. 
The energy scales span several orders of magnitude and relate physical scales which are 
typically treated independent of each other. The pivotal guiding principle of this work is \textit{classical} scale invariance for which there are manifold motivations.

The starting point is the formulation of our model in the Jordan frame, consistent with gauge symmetries, global scale invariance, and general diffeomorphism invariance. 
To achieve the Coleman-Weinberg-type breaking of scale invariance, the SM is extended by two additional scalars with negligible couplings to the Higgs boson (and thus also to the SM). 
The study of the radiatively generated minimum is carried out by resorting to the Gildener-Weinberg approach which is based on the existence of a flat direction along which 1-loop quantum fluctuations 
induce a finite VEV $v_S$ for the scalar $S$ while the second scalar $\sigma$ can subsequently be integrated out due to its high mass. 
The VEV $v_S$ breaks scale invariance spontaneously which gives rise to both the Planck scale $M_\mathrm{Pl} \approx \beta_S^{1/2} v_s$, as well as to right-handed neutrino masses $m_N = y_M v_S$. The latter radiatively generates the Higgs mass term, which in turn, triggers electroweak symmetry breaking while simultaneously generating 
the light neutrino masses via the type-I seesaw mechanism; a scenario dubbed the ``neutrino option''. 
For this process to work, we require right-handed neutrino masses of order $m_N \sim 10^7$~GeV, which can be obtained if the Yukawa coupling to the scalar $S$ 
is of the order $y_M \sim 10^{-(9-10)}$ -- a ``smallness'' that is technically natural. 
In addition, the Higgs portal $\lambda_{HS}$ must be tuned to a small value in order to avoid generating a large Higgs mass term directly from $v_S$. 
This fine-tuning is not technically natural but we have argued that it is not spoiled by quantum corrections if $y_M$ is small enough.

Once a finite scale has been generated, one can perform a Weyl rescaling of the metric to transform to the Einstein frame. 
In this frame, the gravity sector is described by the sum of the Einstein-Hilbert action and non-minimal coupling interactions 
which are transmitted to an involved scalar potential that we use to realize cosmic inflation consistent with observational data. 
Including the globally scale invariant $R^2$-term effectively yields a new scalar degree of freedom, the scalaron, yielding a two-field scalar potential in the Einstein frame. 
A general feature of this potential, related to the absence of an explicit mass scale, is the existence of a flat valley structure in field space 
along which the slow-roll conditions of inflation are satisfied.  
Assuming that the inflationary slow-roll trajectory is confined to this valley (which is justified by a detailed analysis presented in appendix~\ref{app:valley}), 
we have used an effective one-field inflaton potential to simplify the study of inflation and predict CMB observables.
We have performed a parameter scan resulting in values of the scalar spectral index in the range $0.964 \lesssim n_s \lesssim 0.975$ and tensor-to-scalar ratio 
$r \lesssim 0.08$, see Fig.~\ref{ns-r-prediction}. 
The inflaton potential considered here is, therefore, consistent with the tightest observational constraints of the Planck collaboration.

Devoid of an explicit reheating mechanism, we can, given the curvature 
of the potential around the minimum and assuming the number of e-folds between horizon crossing of the pivot scale to the end of inflation to be in the range $50 < N_e < 60$, estimate a bound on the reheating temperature. Given these bounds, we also estimate the production of particle dark matter during reheating. To this end, we include an additional set of right-handed $Z_2$-odd Majorana neutrinos $\chi$ that do not participate in the neutrino option. The dominant production process is identified as the direct decay of the inflaton $S \to \chi \chi$, and the correct relic abundance is obtained for a dark matter mass ranging from $10^4-10^8$\,GeV, which is in turn directly related to the reheating temperature $T_\mathrm{RH}$, which varies between $10^{3}-10^{17}$\,GeV, see Fig.~\ref{TRH-DM}.

Our discussion may be extended in the following directions: (i) The scale anomaly, which is responsible for the spontaneous symmetry breaking of scale invariance, also generates a finite zero-point energy that is much larger than the observed cosmological constant. We set aside this problem in this work, arguing that the cosmological constant might be matched by other sectors. Eventually, a model should be formulated in which the scale anomalies from different sectors are matched explicitly. 
(ii) To derive the Coleman-Weinberg potential, we only considered 1-loop fluctuations due to the matter sector. 
A next step would be to also include gravitational 1-loop corrections to the scalar potential (see, e.g.\ Ref.~\cite{Vicentini:2019etr}), and account for any changes brought on by the inclusion of a non-zero frame discriminant after transforming to the Einstein frame \cite{Falls:2018olk}. 
(iii) As our construction is based on scale invariance, the Weyl tensor squared term should not be omitted in a complete model. 
We have assumed this term to be negligible to study inflation, while it is generally known that this term is problematic because it gives rise to a spin-2 ghost. 
This issue might be remedied, for instance, by using the recently introduced fakeon-prescription~\cite{Anselmi:2018ibi}, where the effect on inflation is investigated in Ref.~\cite{Anselmi:2020lpp}. See also \cite{Eliezer:1989cr,Jaen:1986iz,Simon:1990ic,Biswas:2011ar} and the discussion in footnote~1.
(iv) Furthermore, we have assumed throughout that the breaking of the global scale symmetry and the global dark-matter-stabilizing $Z_2$ symmetry 
by non-perturbative effects in quantum gravity~\cite{Abbott:1989jw,Kallosh:1995hi} (see also \cite{Harlow:2018tng} for an investigation based on AdS/CFT correspondence) is negligibly small. Indeed, a non-negligible breaking of scale invariance, for example a mass term $\Lambda^2 S^2$ in the effective potential (\ref{Ueff1}),
would considerably change our predictions of inflationary parameters.
It will be a challenging future task to estimate such quantum effects in a classically scale invariant model like ours.\footnote{
However, note that neglecting effects of a potential breaking of the global scale symmetry seem well-justified given the presence of the conformal anomaly, 
which gives small contributions to the gravitational field equations that lead to the absence of the problematic wormhole solutions to the gravitational equations of motion as demonstrated in~\cite{Kallosh:1995hi}.}
Ultimately, hence, one would like to extend our framework and promote scale invariance to a local symmetry which would not be susceptible to breaking by quantum gravity effects 
and explore the alterations of predictions as compared to the present study.
(v) Finally, this work could be extended by a full two-field study of inflation and its effect on primordial non-Gaussianities (see, e.g.\ Ref.~\cite{Wands:2007bd}). These additional topics aside, we hope that the work presented here may propagate new ideas for dynamical generation of all scales in Nature and, in particular, for the interplay of scale generation and cosmic inflation.

\appendix
\section{Discussion on the valley approximation}
\label{app:valley}
\begin{figure}[h]
	\begin{center}
		\includegraphics[width=6in]{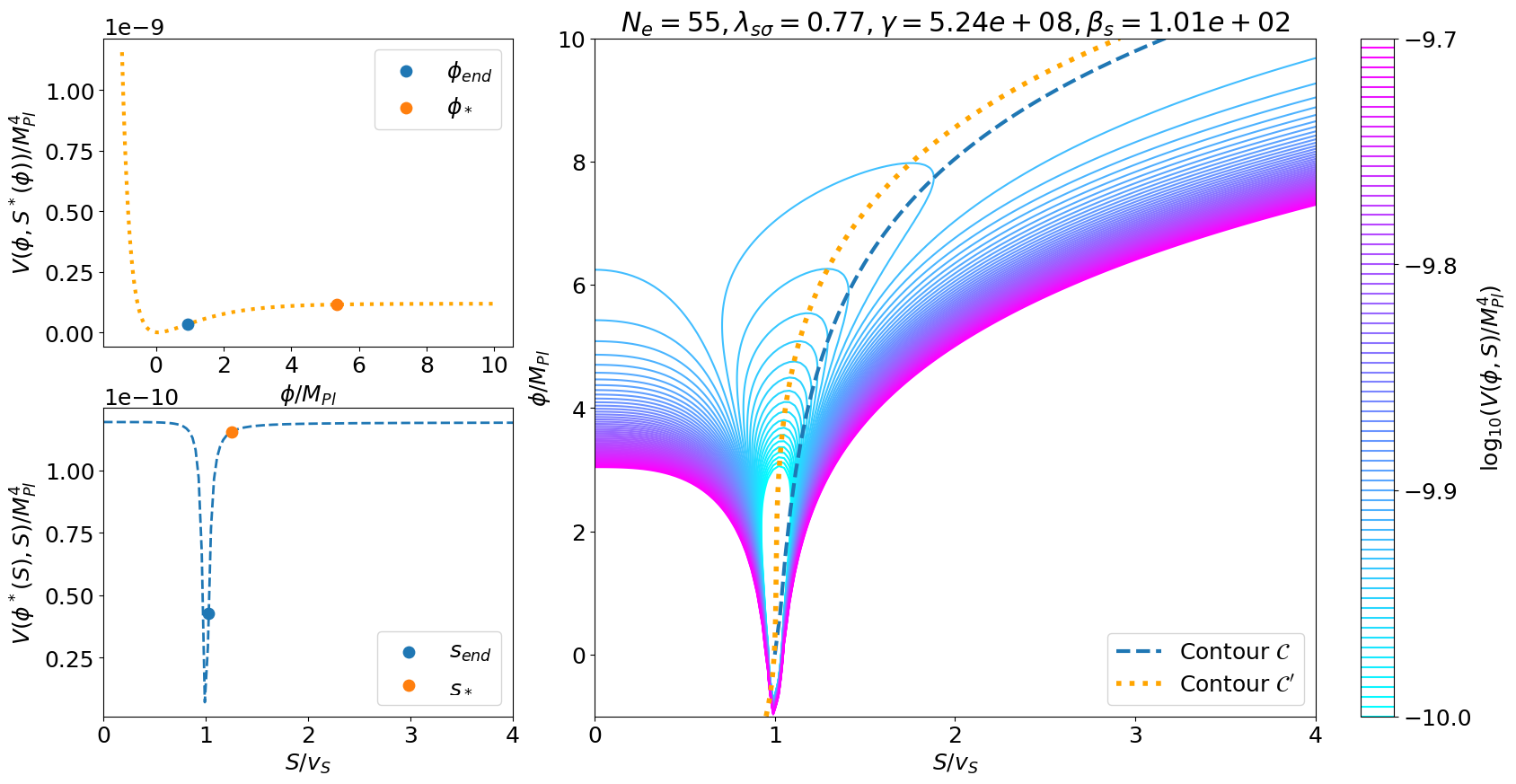}
		\vspace{-2mm}
		\caption{\footnotesize{Scalar potential for benchmark point 1 defined in table~\ref{tablebp}. The two contours defined in Eqs.~\eqref{contourc} and \eqref{contourcprime} are shown on top of the contour plot of $V(S,\phi)$ \eqref{VSphi} (right) and the corresponding 1d inflaton potentials (left).
		}}
		\label{contour_bp0}
	\end{center}
\end{figure}

\begin{figure}[h]
	\begin{center}
		\includegraphics[width=6in]{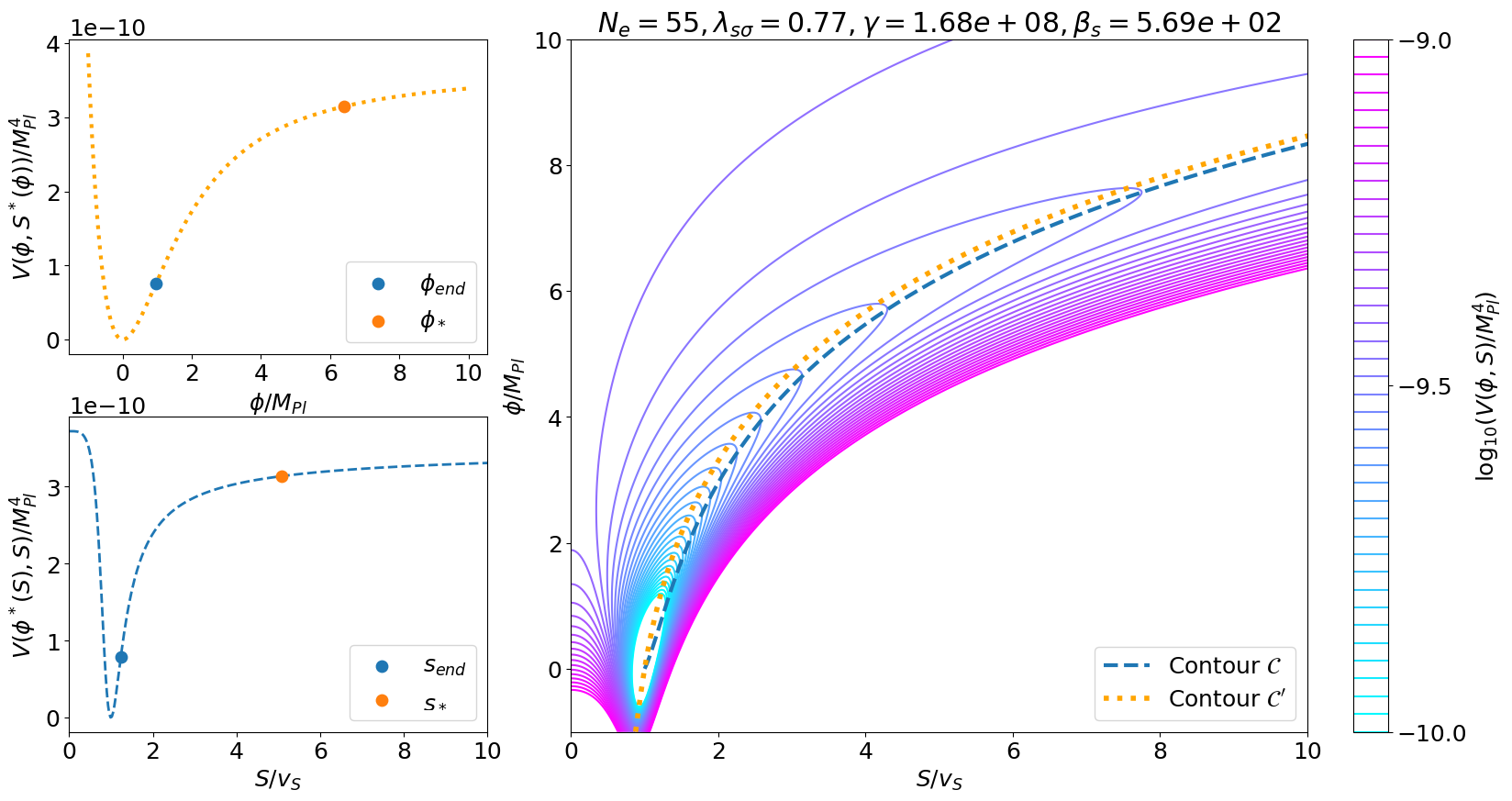}
		\vspace{-2mm}
		\caption{\footnotesize{Scalar potential for benchmark point 2 defined in table~\ref{tablebp}. The two contours defined in Eqs.~\eqref{contourc} and \eqref{contourcprime} are shown on top of the contour plot of $V(S,\phi)$ \eqref{VSphi} (right) and the corresponding 1d inflaton potentials (left).}
		}
		\label{contour_bp1}
	\end{center}
\end{figure}

\begin{figure}[h]
	\begin{center}
		\includegraphics[width=6in]{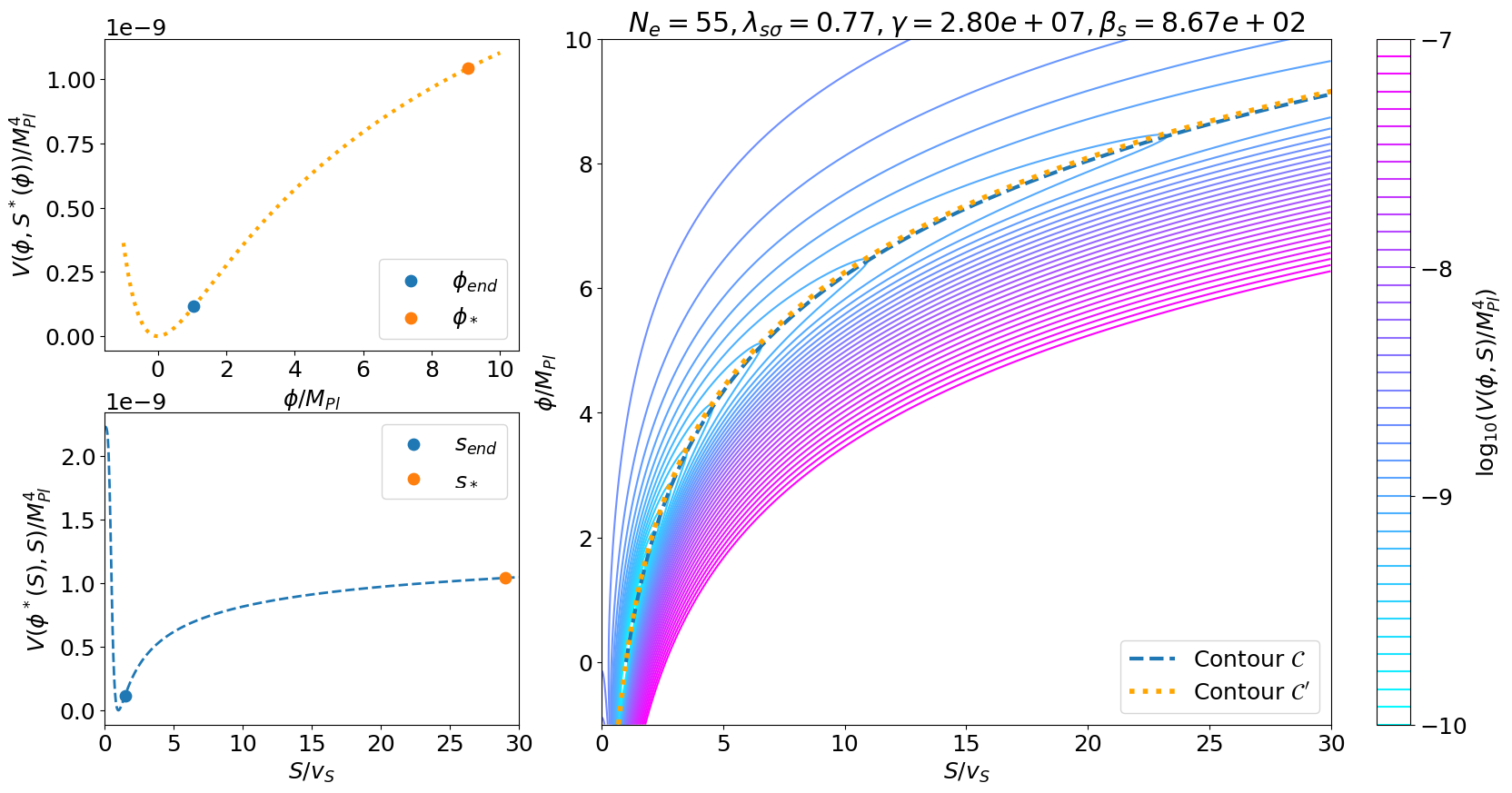}
		\vspace{-2mm}
		\caption{\footnotesize{Scalar potential for benchmark point 3 defined in table~\ref{tablebp}. The two contours defined in Eqs.~\eqref{contourc} and \eqref{contourcprime} are shown on top of the contour plot of $V(S,\phi)$ \eqref{VSphi} (right) and the corresponding 1d inflaton potentials (left).}
		}
		\label{contour_bp2}
	\end{center}
\end{figure}

\begin{figure}[h]
	\begin{center}
		\includegraphics[width=0.65\textwidth]{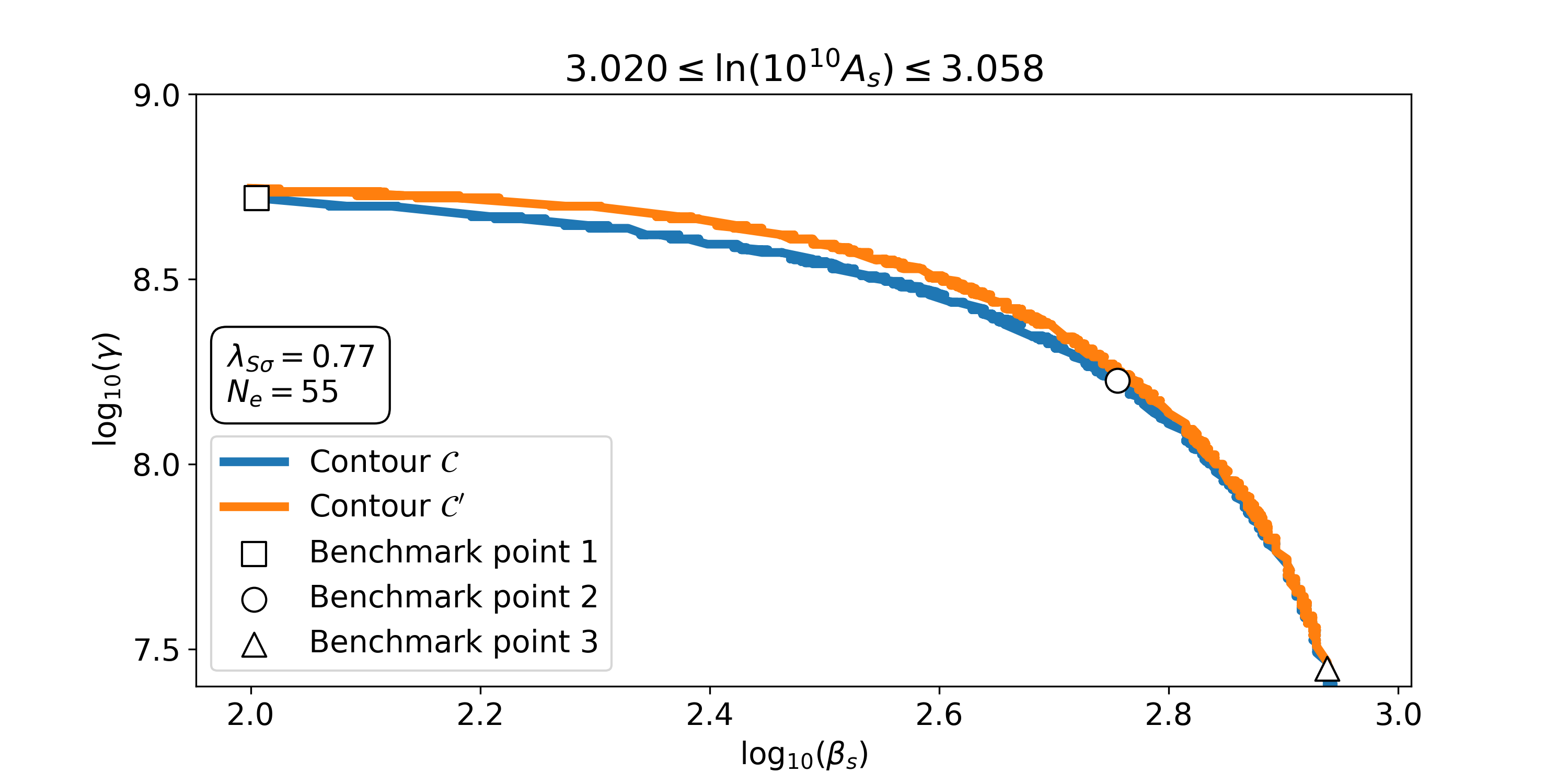}
		\vspace{-2mm}
		\caption{\footnotesize{The lines indicate the parameter combinations of $\gamma$ and $\beta_S$ for which the scalar power spectrum amplitude $A_s$ prediction is fixed to the Planck constraint \eqref{As_constraint} for the two inflationary contours defined in Eqs. \eqref{contourc} and \eqref{contourcprime}. For all points we have fixed $\beta_\sigma =1$ and $\lambda_S = 0.005$. The three benchmark points defined in table~\ref{tablebp} and displayed in Fig.~\ref{contour_bp0} - \ref{contour_bp2} are marked.}
		}
		\label{As_comparison}
	\end{center}
\end{figure}

In this appendix we return to the task of determining the valley contour of the effective potential in the two-dimensional field space. In particular, we compare two different methods for obtaining the valley contour numerically for different parameters of the model. This corroborates the discussion in section \ref{sec:valley_approx} and justifies the method used in section \ref{sec:pscan}.
The method used so far is based on searching for minima in the $\phi$-direction, yielding the contour $\mathcal{C}$ \eqref{contourc}. Alternatively, one can compute the minima along the $S$ direction to define the contour $\mathcal{C}^\prime$ \eqref{contourcprime}.
In the latter case we have to solve for $\tilde S(\phi)$ numerically by finding the minimum for each value of $\phi$ in the $S$-direction. To this end we fit to a seventh order polynomial and yielding the one-dimensional inflationary potential parameterized by the scalaron $\phi$. Thus, the inflaton field is identified with $\phi$ and the slow-roll parameters and number of e-folds are defined accordingly (cf. Eqs. \eqref{epsilon} - \eqref{efolding}). 
To exemplify the two mentioned methods above, we show contour plots of the two-dimensional potential including the contours specifying the two valley approximations  in Fig.~\ref{contour_bp0} - \ref{contour_bp2} for the three benchmark points defined in table~\ref{tablebp} (see also Fig.~\ref{As_comparison}).


On the left hand side of these figures we show the one-dimensional inflationary potentials which correspond to the contours $\mathcal{C}$ and $\mathcal{C}^\prime$. In the case of $\mathcal{C}$, 
the free variable of the potential is $S$, while in the case of $\mathcal{C}^\prime$ the variable is $\phi$. Alternatively, we could, for example, invert $\tilde{S}(\phi)  = S$ to make both one-dimensional potentials dependent on $S$ and then directly compare them. However, as the inflation predictions are computed with $V_\mathrm{inf}(S) (V_\mathrm{inf}(\phi))$ for contour $\mathcal{C} (\mathcal{C}^\prime)$ we have chosen to plot parameterizations as given.
Comparing the three contour plots shows that for lower value of $\gamma$ the valley extends more in the $S$-direction, 
indicating that contour $\mathcal{C}$ is the better choice. However, both methods give very similar results (see e.g.\ Fig.~\ref{contour_bp2}) and approximate the valley contour well. 
For even lower values for $\gamma$ than in benchmark point 3, we run into numerical problems using the contour $\mathcal{C}^\prime$, indicating that the lowest value where we can test both contours is $\gamma \sim 10^7$. 
The two contours deviate more from each other as the parameter $\gamma$ grows large. For large $\gamma$ the inflation field becomes better identified with the scalaron $\phi$ and the valley points more in that direction. Here the two contours start to deviate and contour $\mathcal{C}^\prime$ is expected to be the better approximation. 

\begin{figure}[h]
	\begin{center}
		\includegraphics[width=.9\textwidth]{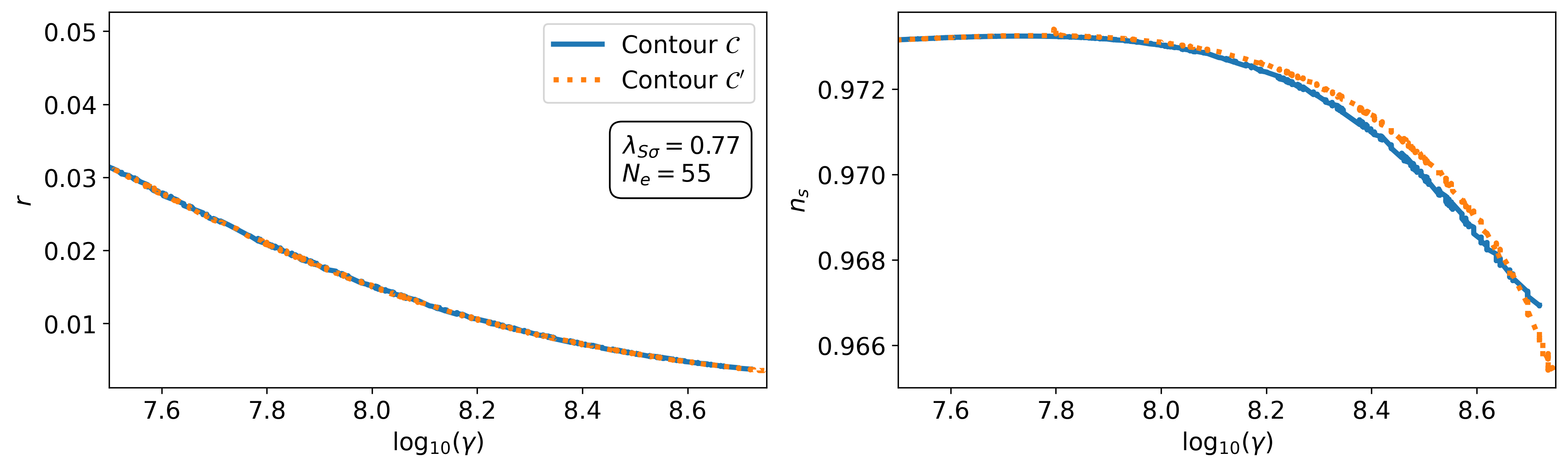} 
		\vspace{-2mm}
		\caption{\footnotesize{Inflation parameters computed along the two contours $\mathcal{C}$ and $\mathcal{C}^\prime$: Tensor-to-scalar ratio $r$~(left) and scalar spectral index $n_s$~(right). Note that the relation between $\gamma$ and $\beta_S$ is different for contours 
		$\mathcal{C}$ and $\mathcal{C}^\prime$, which can be understood from Fig.~\ref{As_comparison}.
			}
		}
		\label{ns_and_r_comparison}
	\end{center}
\end{figure}

A more quantitative comparison indeed shows that the difference of the two contours at large $\gamma$ leads to different predictions for the CMB observables. Fig.~\ref{As_comparison} shows how the experimental constraint on $A_s$ is satisfied for different pairs of $\gamma$ and $\beta_S$ when comparing both methods. Note that this results in different relations between $\gamma$ and $\beta_S$ for both methods which are used in the following.
We use the relations in Fig.~\ref{ns_and_r_comparison} to show the effect on predictions for $n_s$ and $r$ separately, and with varying $\gamma$. The figures show that it is mostly the $n_s$ predictions at large $\gamma$ which vary with the contour. The deviation is relatively small, which justifies the use of contour $\mathcal{C}$ for the larger parameter scan in section \ref{sec:pscan}.


\acknowledgments

J.R.\ and P.S.\ are supported by the IMPRS-PTFS.
J.\ Kubo is partially supported by the Grant-in-Aid for Scientific Research (C) from the Japan Society for Promotion of Science (Grant No.19K03844).

\bibliographystyle{JHEP}
\bibliography{uosref1}

\end{document}